\documentclass[12pt,nohyper,letterpaper]{JHEP3}         
\usepackage{amssymb,amsfonts}

\usepackage{epsfig}

\def\be{\begin{eqnarray}}
\def\ee{\end{eqnarray}}
\newcommand{\nn}{\nonumber}
\newcommand\para{\paragraph{}}

\newcommand{\eqn}[1]{(\ref{#1})}

\def\Dslash{\,\,{\raise.15ex\hbox{/}\mkern-12mu D}}
\def\Dbarslash{\,\,{\raise.15ex\hbox{/}\mkern-12mu {\bar D}}}
\def\delslash{\,\,{\raise.15ex\hbox{/}\mkern-9mu \partial}}
\def\delbarslash{\,\,{\raise.15ex\hbox{/}\mkern-9mu {\bar\partial}}}
\def\pslash{\,\,{\raise.15ex\hbox{/}\mkern-9mu p}}
\def\calDslash{\,\,{\raise.15ex\hbox{/}\mkern-12mu {\cal D}}}

\newcommand{\RN}{Reissner-Nordstr\"om\ }

\def\lae{\mathrel{\mathop{\smash{\lower .5 ex \hbox{$\stackrel<\sim$}}}}}
\def\lae{\mathrel{\mathop{\smash{\lower .5 ex \hbox{$\stackrel>\sim$}}}}}

\def\theequation{\thesection.\arabic{equation}}

\title{Monopoles and Holography}
\author{
Stefano Bolognesi and David Tong \\
Department of Applied Mathematics and Theoretical Physics, \\
University of Cambridge, UK\\
{\tt s.bolognesi, d.tong@damtp.cam.ac.uk}
}

\abstract{We present a holographic theory in $AdS_4$ whose zero temperature ground state develops a
crystal structure, spontaneously breaking translational symmetry. The crystal is induced by a
background magnetic field, but requires no chemical potential. This lattice arises from the
existence of 't Hooft-Polyakov monopole solitons in the bulk which condense to form a classical
object known as a monopole wall. In the infra-red, the magnetic field is screened and there is an
emergent $SU(2)$ global symmetry.}

\begin{document}
\pagestyle{plain} \setcounter{page}{1}
\newcounter{bean}
\baselineskip16pt

\section{Introduction}

The AdS/CFT correspondence offers a unique opportunity to explore the possible dynamics of strongly
interacting matter in a controlled setting. While its relevance to any specific system that can
realised in the laboratory is likely tenuous, it nonetheless provides a controlled framework in
which we can ask the simple question: what can strongly interacting matter do?

\para
With this motivation, it is worthwhile to study various phenomena in the bulk to look for novel
physics that can occur in the boundary theory. In this paper we ask what novel physics arises if
the bulk contains dynamical magnetic monopoles.

\para
The dynamical magnetic monopoles that we have in mind live in $AdS_4$. They are not fundamental,
point-like objects since these could be readily understood by performing a bulk S-duality. Instead
we will be interested in the role played by bulk, solitonic, 't Hooft-Polyakov monopoles. The
simplest theory in which such objects appear is an $SU(2)$ gauge theory, spontaneously broken to
$U(1)$ by an adjoint scalar field. This is the same bulk theory considered in \cite{hongmag} as a
model of anti-ferromagnetism, but with different asymptotic behaviour for the fields corresponding
to different sources in the boundary theory.

\para
By the usual holographic dictionary, the boundary conformal field theory enjoys a $U(1)$ global
symmetry. Interesting physics occurs when a background, homogeneous, magnetic field is turned on
for this $U(1)$. The main goal of this paper is to study the ground state of this system. One
candidate ground state is the familiar, magnetically charged, \RN black hole. However, we will show
that at low temperatures this is not the preferred ground state; that honour goes instead to an
object known as a {\it monopole wall}.

\FIGURE{\epsfig{file=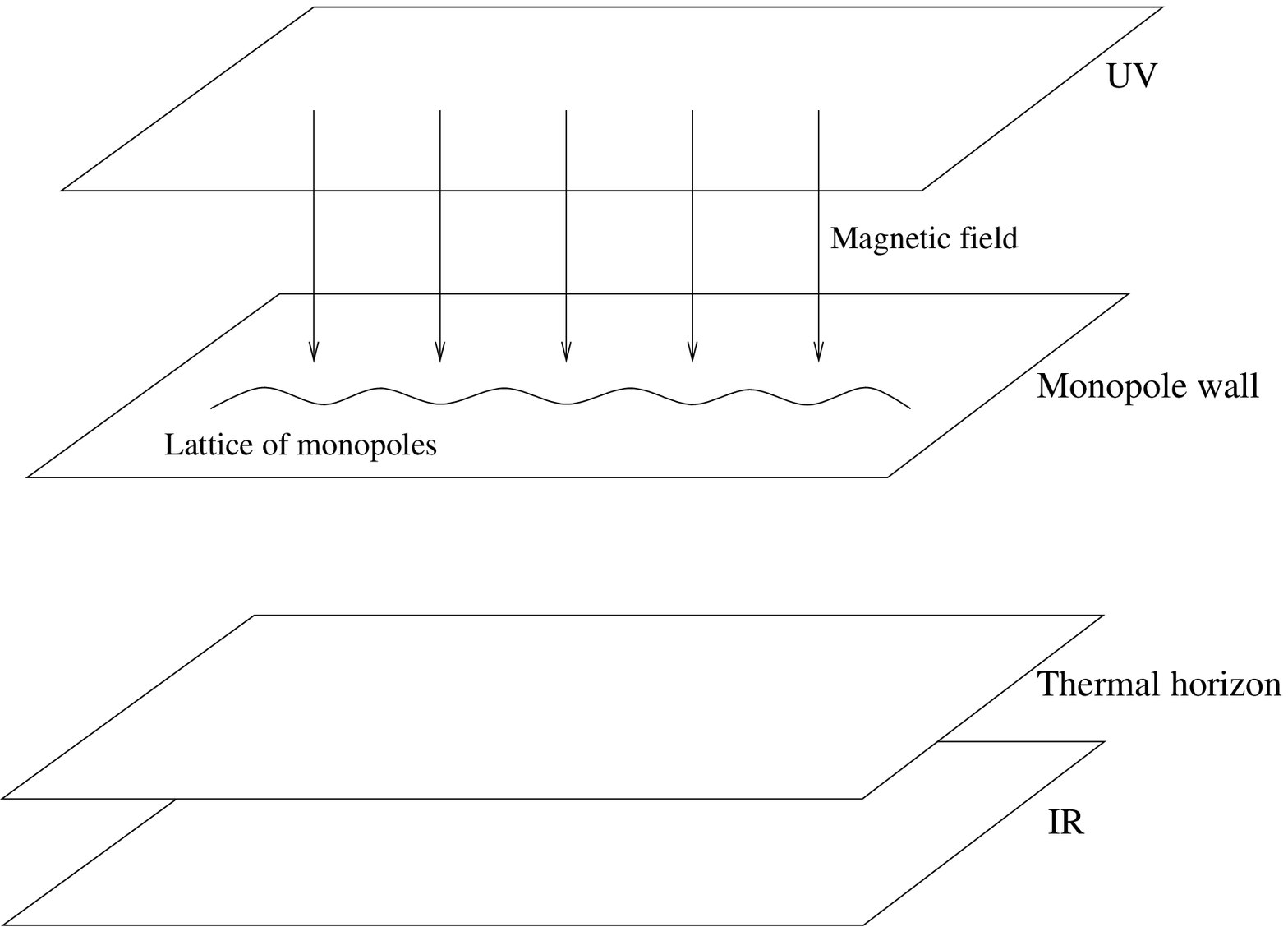,width=8cm}
        \caption[]{Sketch of the monopole wall in AdS}%
	\label{monopolewall}}
\para
The monopole wall is a complicated and poorly understood object. There is currently no known
explicit solution. Various arguments for the existence of the monopole wall were given in
\cite{bag} and a numerical study providing compelling evidence for these solutions was performed by
Ward \cite{ward}. The purpose of this paper is to describe the implications of the existence of the
wall for the boundary field theory.

\para
At the simplest level, the monopole wall should be thought of as domain wall interpolating between
the broken and unbroken phase of the bulk gauge theory. However, unlike common-or-garden domain
walls, it is not translationally invariant. Instead it forms a crystalline lattice structure. In
terms of the AdS/CFT dictionary, the lattice structure of the wall in the bulk is imprinted on the
expectation values of the boundary theory, which spontaneously breaks translational symmetry. The
culprits are operators dual to both the neutral scalar Higgs field and to the W-boson fields. The
latter carries both rotational and $U(1)$ quantum numbers, ensuring that the ultra-violet physics
is a that of a superconductor. However, the physics in the infra-red is very different: the
magnetic field is screened and the bulk gauge symmetry is restored, corresponding to an emergent
$SU(2)$ global symmetry on the boundary. Another interestic development, would be to consider dyonic monopole walls and study the Hall conductivity (see e.g. \cite{Hartnoll:2007ai} for dyonic black holes).

\para
In some sense, our set-up is a magnetic version of the familiar story of holographic
superconductors \cite{gubser,hhh} (see \cite{sean,gary} for nice reviews). In that case, an
electric \RN $AdS$ black hole has a linear instability to the formation of hair if the theory also
includes a suitably charged scalar field. In contrast, the magnetic \RN $AdS$ black hole does not
appear to have a linear instability; rather the true crystal ground state is separated from the
black hole by an energy barrier. It is interesting to speculate that, via S-duality, our weakly
coupled magnetic black hole captures the physics of the electric holographic superconductor when
the bulk gauge coupling is large. Indeed, it was suggested in \cite{sachdev} that the correct
holographic theory of a strange metal consists of spontaneously formed crystal of spins of the type
we observe.

\para
This paper has three further sections. In section 2 we explain how to construct the monopole wall
and describe the asymptotic behaviour of the fields in AdS, corresponding to the sources of the
boundary theory. We focus on an Abelian approximation, first introduced in \cite{bag}. This has the
advantage that we can determine various crude properties of the monopole wall in $AdS$ but the
approximation is too myopic to capture the most interesting feature, namely the lattice structure.
In Section 3, we discuss the physics of the monopole wall seen in the boundary field theory. We
present estimates of the lattice spacing and amplitude imprinted on the expectation values of
boundary operators. In both Sections 2 and 3, we work in a probe approximation in which matter
fields live on a fixed $AdS$ background. In Section 4 we consider the gravitational backreaction of
the monopole wall and the melting of the lattice at finite temperature.

\section{Monopoles in AdS}

In this Section we will describe the monopole wall in $AdS_4$, both in global coordinates and in
the Poincar\'e patch. Our main goal is to examine the field asymptotics that are required to
support the wall. These correspond to sources in the boundary field theory. Importantly, these
sources preserve translational invariance.

\para
Our starting point is the simplest model which contains 't Hooft-Polyakov monopoles: an $SU(2)$
gauge field $A_\mu^a$ and a single adjoint Higgs field $\phi^a$. This is coupled to gravity through
the Einstein-Yang-Mills-Higgs action,
\be S = \int d^4x \sqrt{-g}\,\left[\frac{1}{2\kappa^2}\left(R+\frac{6}{L^2}\right) -
\frac{1}{e^2}\left(\frac{1}{4}F^a_{\mu\nu}F^{a\mu\nu}+\frac{1}{2}{\cal D}_\mu \phi^a {\cal D}^\mu
\phi^a + V(\phi)\right)\right]\label{action}\ee
The potential $V(\phi)$ is given by
\be V(\phi) = \frac{\lambda}{8}(\phi^a\phi^a-v^2)^2 \ee
This same theory was considered in \cite{hongmag} as a model of holographic anti-ferromagnetism,
while bulk $SU(2)$ gauge theories have previously been used to construct models of p-wave
superconductors \cite{pwave1,pwave2}.

\para
We will ask that our theory in this background is weakly coupled at the symmetry breaking scale
$v$. In the vacuum, the Higgs field sits at the minimum of the potential, $\phi^a\phi^a=v^2$,
breaking the $SU(2)$ gauge symmetry to $U(1)$. The background spacetime is $AdS_4$ with the planar
metric
\be ds^2 = \frac{r^2}{L^2}(-dt^2 + dx^2 +dy^2) + \frac{L^2}{r^2}dr^2\ee

\para
The details of the boundary CFT depend on the parameters in the potential. It will be useful to
separate the potential into three classes:
\begin{itemize}
\item $\lambda=0$: When the potential vanishes, the boundary theory has an $SU(2)$ global
    symmetry, with current operators $J_\mu^a$. These have dimension 2, as befits a conserved
    current. The scalar fields are dual to a triplet of marginal operators $\Phi^a$. The
    asymptotic behaviour of $\phi$ is given by
\be \phi^a(r) \rightarrow v^a + \frac{\beta^a}{r^3}\label{phispecial}\ee
    where $v^a$ is a source for the operator $\Phi^a$. We will assume that the magnitude of the
    source, $v^a v^a$, is constant on the boundary. This source {\it explicitly} breaks the
    $SU(2)$ global symmetry of the boundary, and the triplet of currents now obey
\be \partial^\mu J^a_\mu = \epsilon^{abc} v^b {\cal O}^c \ee
    If $v^a$ is constant on the boundary, then $SU(2)$ is broken only to $U(1)$. However, for
    general $v^a(\vec{x})$, it is broken completely. In Section (and, in more detail, in the
    Appendix) we will specify more precisely the conditions under which a global $U(1)$
    survives even in the presence of a spatially varying  $v^a(\vec{x})$.

\item $\lambda<0$: The BF bound allows $\lambda v^2L^2 > -9/4$. In this case, the boundary
    symmetry breaking is not induced by a source. Rather, the bulk theory has only a massless
    $U(1)$ photon, corresponding to a conserved $U(1)$ boundary current $J_\mu$. The W-boson in
    the bulk has mass $v$ and is associated to a charged, spin-1 operator ${\cal W}_\mu$ of
    dimension $\Delta_W = 3/2 + \sqrt{1/4+v^2L^2}$. Meanwhile, the bulk scalar field that isn't
    eaten in the Higgs mechanism is dual to a relevant operator $\Phi$. The asymptotic fall-off
    is given by
\be \phi^a \rightarrow \hat{v}^a\left(v + \frac{\alpha}{r^{\Delta_-}} +
\frac{\beta}{r^{\Delta_+}}\right)\label{phiexp}\ee
with
\be \Delta_\pm = \frac{3}{2}\pm \sqrt{\frac{9}{4} + \lambda v^2}\ee
For $-9/4< \lambda v^2L^2 < -3/2$, there is the usual ambiguity in quantization of this scalar.
In this paper, we choose the standard quantization in which $\alpha$ is interpreted as the
source, and $\beta\sim \langle\Phi\rangle$. The dimension of $\Phi$ is then given by $3/2 <
\Delta_+ < 3$.

\item $\lambda>0$: The discussion is much the same as the $\lambda < 0$ case, except the scalar
    operator $\Phi$ is now irrelevant.
\end{itemize}

Usually in the discussions of holography, one specifies the asymptotic sources and solves the bulk
equations of motion, subject to an appropriate boundary condition in the infra-red. In our case,
the magnetic monopole requires topologically non-trivial winding of the scalar field which results
in some subtleties in the sources. For this reason, we will present the discussion in reverse: we
will firstly describe the magnetic monopole solutions in the bulk of $AdS$ and then examine what
sources are required to support these objects. Ultimately, it will turn out that these sources are
actually very simple, but we will have to work to see this.

\para
For the remainder of this Section, and throughout Section 3, we work in a fixed $AdS_4$ background,
neglecting the backreaction on spacetime. Roughly, this approximation is valid when
$v^2\kappa^2/e^2 \ll 1$ since this ratio governs the relative strength of the gravitational and
electromagnetic forces. However, even in this regime it may be necessary to take backreaction into
account. (For example, a constant magnetic field will always backreact in the far infra-red). In
Section 4, we re-instate Newton's constant and include backreaction to describe transitions between
the monopole wall and the black hole.

\subsection{Monopoles in Global AdS}

The 't Hooft-Polyakov monopole is a solution that owes its existence to the topology of the Higgs
sector. Because the solution requires scalar fields to wind at infinity,  it is instructive to
start our discussion in global $AdS_4$ spacetime which has boundary ${\bf R}\times{\bf S}^2$. The
metric is
\be ds^2 = -\left(1+\frac{\rho^2}{L^2}\right)d\tau^2 +
\left(1+\frac{\rho^2}{L^2}\right)^{-1}d\rho^2 + \rho^2d\Omega_2^2 \label{global}\ee
Asymptotically, as $\rho \rightarrow \infty$, the Higgs field is required to sit in its vacuum
moduli space ${\bf S}^2=\{v^a: v^a v^a =v^2\}$. Maps from the boundary of $AdS_4$ to this vacuum
manifold are labeled by an integer $n\in {\bf Z} \cong \Pi_2({\bf S}^2)$.

\para
In flat spacetime, there are standard energetic arguments which ensure that any winding is
accompanied by a long range magnetic field (see, for example, \cite{harvey}). These arguments also
hold in $AdS_4$. Specifically, to avoid a linear divergence in the energy, the asymptotic gauge
field must turn on to cancel, at  the leading order, $\partial_i \phi^a$. This means that the
covariant derivative
\be {\cal D}_\mu\phi^a \equiv \partial_\mu \phi^a + \epsilon^{abc} A_\mu^b\phi^c =0 \label{dphi}\ee
to order ${\cal O}(1/\rho)$. Solving this equation then guarantees that the long range magnetic
field lives entirely in the unbroken $U(1)$ subgroup,
\be F^a_{\mu\nu} = -\frac{\,\phi^a}{v} F_{\mu\nu}\nn\ee
where
\be F_{\mu\nu} = \frac{1}{v^3}\epsilon^{abc}\phi^a\partial_\mu\phi^b\partial_\nu\phi^c
-\partial_\mu C_\nu+\partial_\nu C_\mu\label{longmag}\ee
for some (globally defined) Abelian gauge field $C_\mu$
%
%
%
%
%

\para
Let's discuss the necessary boundary conditions to build a monopole, focussing first on the case
with winding $n=1$\footnote{Although tangential to the main topic of the paper, one may wonder
about the connection between 't Hooft-Polyakov monopoles in the bulk and the objects in 3d CFTs
that are usually referred to as monopole operators (see, for example, \cite{kapustin} for a modern
perspective). The connection is made if we choose the alternative quantization for the $U(1)$ gauge
field \cite{witten,ross}, effectively gauging the $U(1)$ global symmetry of the boundary CFT. The
dimension of the monopole operator is then equal to the lowest energy state which carries unit
magnetic charge. One candidate for such an operator is the extremal Reissner-Nordstr\"om black
hole, corresponding to an operator of dimension $\sim 1/\kappa^2$. The 't Hooft-Polyakov monopole
provides a different state, carrying the same quantum numbers, but with parameterically lower
energy.}. Smoothness requires that $\phi=0$ at a point in the interior of $AdS_4$ which
we take to be $\rho=0$. We also require that the Higgs expectation value winds once at infinity;
the standard choice is the hedgehog ansatz $v^a = v \hat{x}^a$, with $\hat{x}^a$ the unit normal to
the boundary. For $\lambda=0$, the expectation value $v^a$ is the source for $\phi$ and this is
sufficient information to determine the solution  A study of monopoles in global $AdS_4$ in
hedgehog gauge, primarily with $\lambda =0$, was performed some time ago in
\cite{schap,schap2}.  Multimonopoles with axial symmetry in $AdS_4$ where considered in \cite{Radu:2004ys}. Dyons in global AdS where considered in \cite{Allahbakhshi:2010ii}. For $\lambda\neq 0$, the boundary expansion \eqn{phiexp} (suitably generalized to global,
rather than planar $AdS_4$) means that we still have to specify one further boundary condition,
which we take to be $\alpha=0$. (In fact, for $\lambda >0$ this is implicit in the requirement that
$\phi \rightarrow v$ on the boundary, the $\alpha$ mode is a non-normalizable growing mode). We
note that for $-9/4< \lambda v^2L^2 < -3/2$, there should be two monopole solutions: one with
$\alpha =0$ on the boundary and one with $\beta = 0$ on the boundary.

%
%
%

\subsubsection*{Monopole Bags}

In this paper, our interest lies with multi-monopoles. In flat space, static solutions with winding
number $n\geq 2$ only exist in the BPS limit of vanishing potential $\lambda =0$. These are the
well-studied solutions to the Bogomolnyi equations $B_i={\cal D}_i\phi$. When $\lambda
>0$, the magnetic force always beats the (now exponentially suppressed) scalar attraction and
monopole repel.

\FIGURE{\epsfig{file=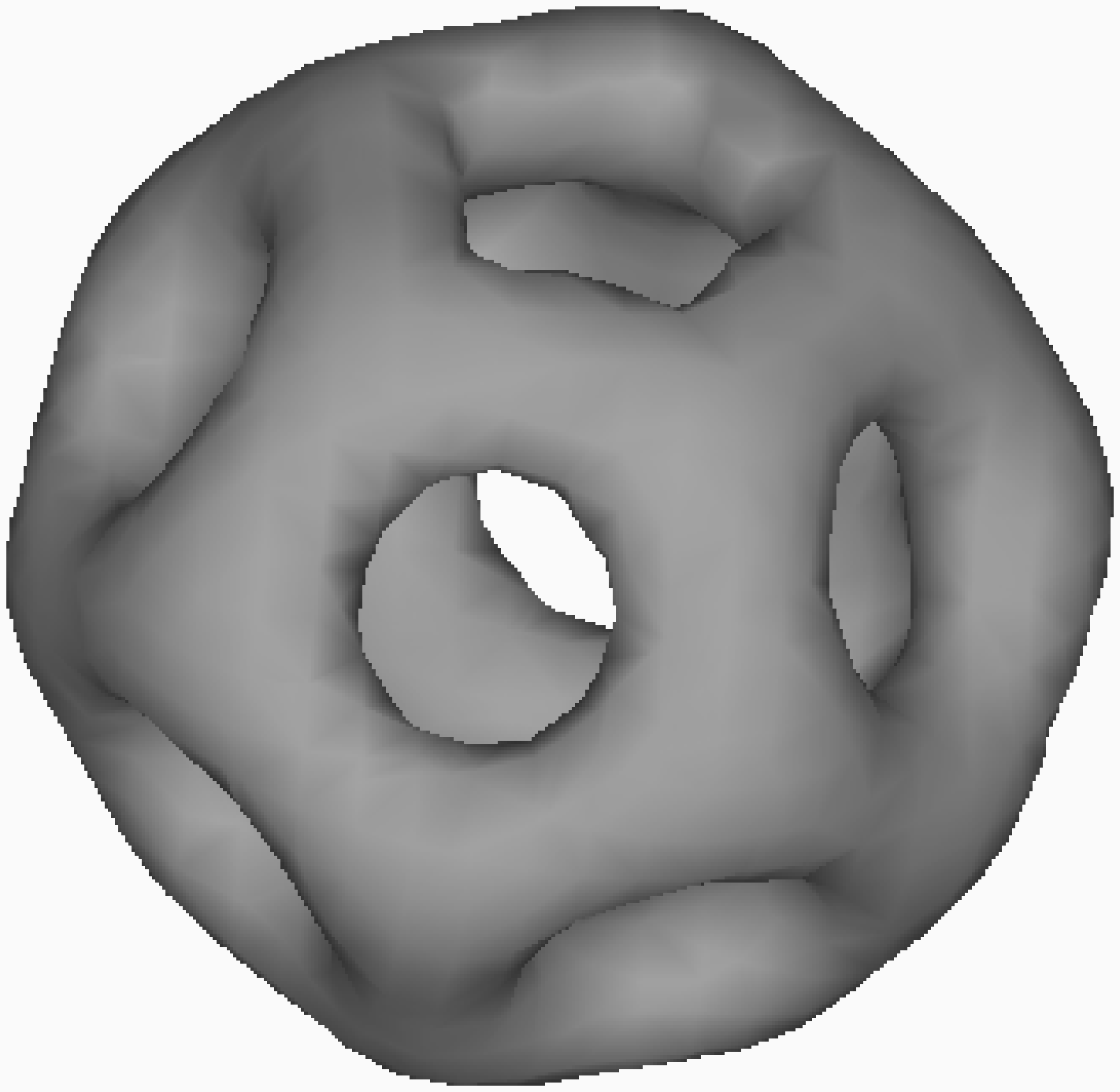,width=5cm}
        \caption[]{The $n=7$ dodecahedron monopole \cite{sm2}}
	\label{stolen}}
\para
In global $AdS_4$ however, the situation is  somewhat richer. First, even for $\lambda >0$, there
remains a gravitational attraction which, at least for small values of $\lambda$, would appear to
allow classical monopole bound states. Furthermore, in the regime $\lambda<0$, the long range
scalar attraction beats the magnetic repulsion and, once again, multi-monopole configurations are
allowed.

\para
To get an intuition for the multi-monopole solutions, let's start by recalling what happens when we
bring BPS monopoles close together in flat space.  Importantly, in this limit monopoles do not
behave like hard-core tennis balls. Instead the non-linear nature of the solutions becomes
important and the monopole core grows to a size

\be R \sim \frac{n}{ v}\label{flatbag}\ee
However, the exact solution cannot be spherically symmetric \cite{guth}. For low winding numbers,
it is known that there exist solutions with toroidal ($n=2$), tetrahedral ($n=3$), octahedral
($n=4$) and icosahedral ($n=7$) symmetry \cite{symmon,sm2}. A plot of the constant energy surfaces
for the $n=7$ dodecahedron monopole is shown in Figure \ref{stolen}, taken from \cite{sm2}. For
higher winding numbers, less is known analytically but it is clear that $n$ monopoles form a large
ball of radius $R$, with a lattice structure on the surface.

\FIGURE{\epsfig{file=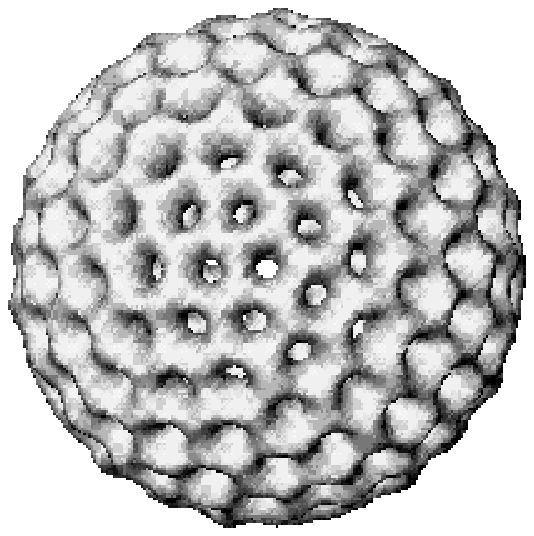,width=5cm}
        \caption[]{The $n=97$ skyrmion \cite{stolen2}}
	\label{stolen2}}
\para
For monopoles in flat space, the exact form of the surface lattice depends on where one sits in
moduli space. In global $AdS$, this degeneracy will be lifted by the gravitational attraction. This
is somewhat analogous to the situation with Skyrmions, where there is again an attraction between
far separated solitons, now due to pion exchange. Skyrmions have a close relationship to magnetic
monopoles \cite{monskyrme}, but the absence of a moduli space (and the absence of gauge symmetry)
means that it is easier to find numerical solutions. Skyrmions with large winding number are known
to form buckball-like objects with a hexagonal lattice (for multiple layers the lattice may be different \cite{SilvaLobo:2009qw}). For example, Figure \ref{stolen2}, taken from \cite{stolen2}, shows a Skyrmion with
winding number $n=97$ with icosahedral symmetry.


\para
While the exact solution is complicated, an approximation was developed in \cite{bag} which
captures the crude features of monopoles with large winding.  This approximation ignores both the
fine structure of the monopole solution, including the lattice and non-Abelian fields. Instead, one
focusses just on the Abelian photon and the scalar field $\phi$ and thinks of the monopole in terms
of a bag model. In flat space, one of the surprises of \cite{bag} was that the BPS bag could take
any shape, reflecting the moduli space of BPS monopoles. In contrast, we are interested in bags in
$AdS_4$ where the gravitational attraction will ensure that the bag is spherical with radius $R$
which will be determined dynamically.

\para
Inside the bag, both the magnetic field and the scalar vanish,
\be \phi = 0 \ \ \ \ ,\ \ \ \ F_{\mu\nu}=0\ \ \ \ \ {\rm for}\ \rho<R\nn\ee
Outside, the bag acts as a source for Abelian magnetic flux,
\be B = \frac{n}{ \rho^2}\ \ \ \ \ {\rm for}\ \rho>R\nn\ee
Meanwhile, the scalar field obeys the equations of motion, interpolating between $\phi=0$ on the
surface of the bag, $\rho = R$ and $\phi=v$ as $\rho \rightarrow \infty$. The explicit profile
depends on the potential $V(\phi)$. When $V(\phi)=0$, it is a simple matter to solve the  equations
of motion to find
\be \phi(\rho) = v\left(\frac{h(R)-h(\rho)}{h(R)-\pi/2}\right)\nn\ee
where the profile function is
\be h(\rho) = \tan^{-1}(\rho/L)+(L/\rho)\nn\ee
For $\rho \ll L$, this profile function looks like $h(\rho)\sim L/\rho$. Meanwhile, for $\rho\gg
L$, we have $h(\rho) \sim \pi/2 +L^3/3\rho^3$.

\para
We still need to determine the size $R$ of the bag. This is dictated by minimizing the
energy\footnote{This is the conserved energy appropriate to $AdS$. The presence of the timelike
Killing vector $\partial_t$ ensures that $J_\mu = T^t_{\ \mu}$ is conserved and the energy is the
integral over the spacelike slice, $E=\int{}^*j = \int \sqrt{g}\,T^t_{\ t}$.},
\be E(R) &=&  \frac{2\pi}{e^2}\int_R^\infty d\rho \, \rho^2\left[ B^2 + (\partial_\rho\phi)^2
(1+\rho^2/L^2)\right] \nn\\ &=& \frac{2\pi}{e^2} \left[\frac{n^2}{R} +
\frac{Lv^2}{h(R)-\pi/2}\right] \ee
Here one sees the two competing effects: the magnetic flux tries to force the bag to grow while the
gradient energy of the scalar field tends to make the bag shrink. The bag radius is determined by
minimizing $E(R)$. There are two interesting limits. When $n \ll vL$, the monopole does not yet
feel the curvature of $AdS_4$ and the size of the bag is given by the flat space result
\eqn{flatbag}. In contrast, when $n \gg vL$, the size of the monopole bag is greater than the AdS
curvature. Here we find the scaling
\be R = \sqrt{\frac{nL}{3 v}} \gg L \label{rbag}\ee
The energy scales as $E\sim \sqrt{n^3 v /Le^4}$.

\para
In flat space, the bag radius \eqn{flatbag} means the the flux per unit area decreases as $1/n$ as
the number of monopoles increases. In contrast, in $AdS$, the flux per unit area approaches a
constant, $\sim 3v/L$ as the number of monopoles increases. This will be important when we discuss
the monopole wall.

\para
Finally, let us look at the necessary conditions to consistently neglect the gravitational
backreaction. We will do a more thorough job in Section 4, but for now we can simply compare the
radius of the extremal black hole to that of the monopole bag. The former again has a radial
magnetic field, $B=n/\rho^2$ for $\rho> R_h$, the horizon. Like the monopole bags, black holes in
global $AdS$ come in two sizes: big and small. The small black holes, with $R_h \ll L$, essentially
have the features of flat space. Ignoring factors of unity, the size of the horizon is
\be R_h \sim \frac{n\kappa}{e}\ \ \ \ \ \ \ \ \ \ \ R_h \ll L\nn\ee
Meanwhile, the large black holes have horizon given by
\be R_h \sim \sqrt{\frac{n\kappa L}{e}}\ \ \ \ \ \ \ \ \ \ \ R_h \gg L\nn\ee
Comparing to the horizon to the radius of the monopole bag, \eqn{flatbag} and \eqn{rbag}, we see
that both scale in the same way with $n$ and $L$. The monopole bag is well outside the horizon of
the black hole when $v^2\kappa^2/e^2 \ll 1$. For the remainder of this section and the next, we
will assume that we satisfy this criterion.

\subsection{Monopole Walls}

The discussion above was in the context of global $AdS$. We now follow the fate of the monopole bag
as we move to the Poincar\'e patch. We will show that, with a suitable scaling, the monopole bags
turn into planar monopole walls which can be thought of as the skin of the bag. Our discussion is
entirely equivalent to that of large black holes in global $AdS$ which turn into planar black holes
in Poincar\'e coordinates.

\para
The Poincar\'e patch of $AdS$ can be thought of as zooming in to a specific region of the global
space. Starting from the metric \eqn{global}, we write the metric on ${\bf S}^2$ as
\be d\Omega_2^2 = \left(\frac{d\chi^2}{1-\chi^2}+\chi^2d\theta^2\right)\ee
define rescaled coordinates
\be \rho = \zeta r\ \ \ ,\ \ \ \tau = \frac{t}{\zeta}\ \ \ ,\ \ \ \chi = \frac{u}{\zeta
L}\label{scaling}\ee
Sending $\zeta \rightarrow \infty$, keeping $r$, $t$ and $u$ fixed, gives the familiar metric on
the Poincar\'e patch of $AdS_4$, with $u$ a radial coordinate in the $x$-$y$ plane,
\be ds^2 = \frac{r^2}{L^2}(-dt^2 + du^2 + u^2d\theta^2) + \frac{L^2}{r^2}dr^2\label{ads}\ee
If the monopole bag is to keep up with this rescaling, we need to place $n\sim \zeta^2$ monopoles
in global $AdS$ so that the wall of the bag lies at a constant position. However, as stressed in
the previous section, the flux per unit area emitted by the wall remains constant in this limit.

\para
The upshot of this rescaling is a planar monopole wall in the Poincar\'e patch of $AdS_4$. In fact,
it's rather easier to describe this wall than the spherical monopole bag in global $AdS$. (In some
sense, our digression into global coordinates was merely to motivate the existence of the wall). We
will now discuss some of the properties of this wall.

\subsubsection*{The Position of the Wall}

We work again in the Abelian approximation. To infra-red side of the wall, $r<R$, the Abelian
fields are vanishing: $\phi=B=0$. To the ultra-violet side of the wall, $r>R$, the magnetic field
is constant while the form of $\phi$ again depends on the potential.

\para
In the case of vanishing potential, $\lambda =0$, one can solve the equations of motion
analytically. The fields are given by
\be \phi = v\left(1-\frac{R^3}{r^3}\right) \ \ \ \ ,\ \ \ \ B={\rm constant}\label{l0prof}\ee

\para
The energy density of the planar wall is given by
\be E(R) = \frac{1}{2e^2}\int_R^\infty dr\ \frac{r^2}{L^2}\left( \frac{L^4}{r^4}B^2 +
\frac{r^2}{L^2}(\partial_r \phi)^2\right) \nn\ee
Once again, the magnetic flux causes the wall to expand towards the ultra-violet, while the
gradient energy causes it to contract towards the infra-red. The radial position of the wall is
then determined dynamically to be
\be R = \sqrt{\frac{ B L^3}{3v}}\ \ \ \ \ \ \ \ \ \ \ \ \ \ \ \ \ \ \ \ \ \ \ (\lambda
=0)\label{r0}\ee
We can use this result to estimate the local flux emitted by the monopole wall. This is related to
the magnetic field $B$ by the $AdS$ warp factor,
\be B_{\rm local} = \left(\frac{L}{R}\right)^2 B = \frac{3v}{L}\label{lflux}\ee
This is the same flux that we saw was emitted from the monopole in global $AdS$.

\para
For $\lambda\neq 0$, the position of the wall deviates from \eqn{r0} but, as we now show, the
dependence on $\lambda$ is slight. The profile for $\phi$ can be determined by solving the equation
of motion numerically, subject to the boundary conditions \eqn{phiexp}. We set the source to
vanish, $\alpha=0$, while $\beta$ is fixed by the requirement that $\phi(R)=0$. Plots of the
profile of the Higgs field are shown in Figure \ref{lambdaradius}.

\para
To determine the position, $R$, of the wall, we must minimize the energy density,
\be E(R)= \frac{1}{2 e^2}\left[\int_R^\infty dr\ \frac{r^2}{L^2}\left( \frac{L^4}{r^4}B^2 +
\frac{r^2}{L^2}(\partial_r \phi)^2 + \frac{\lambda}{4}(\phi^2-v^2)^2\right)+  \int_0^R dr\
\frac{r^2}{L^2} \frac{\lambda v^4}{4}\right]\ \ \ \ \ \ \ \label{lame}\ee
Importantly, the equation of motion for $\phi$ enjoys a scaling symmetry such that if if $\phi(r)$
is a solution then  $\phi(\alpha r)$ is also a solution for any constant  $\alpha$. This has the
effect that the position $R$ of the wall depends only on the dimensionless ratio $\lambda v^2L^2$,
and not on $\lambda$ and $vL$ separately. To see this, notice that we can rewrite the energy as
\be E(R) = \frac{1}{2 e^2} \left( \frac{L^2 B^2}{R} + \frac{R^3 v^2}{L^4} f(\lambda v^2 L^2) + R^3
\frac{\lambda v^4}{12 L^2} \right) \nn\ee
\DOUBLEFIGURE[]{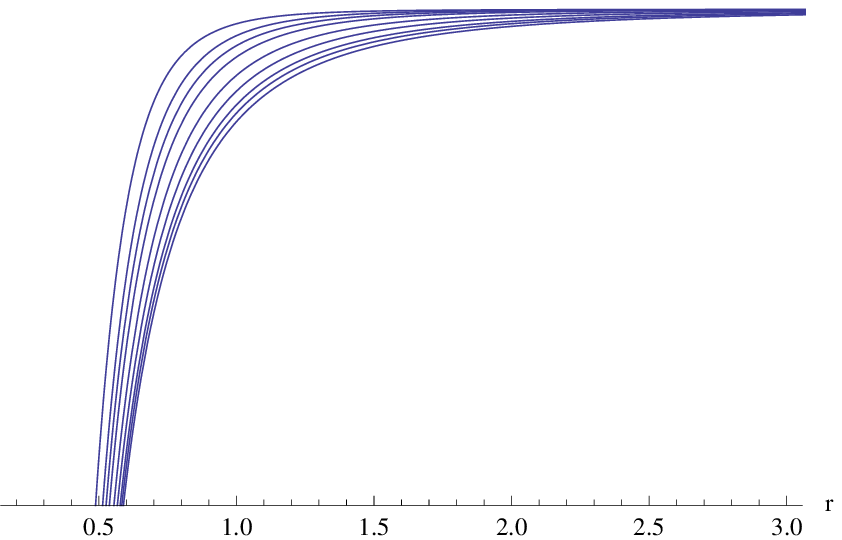,width=7cm}{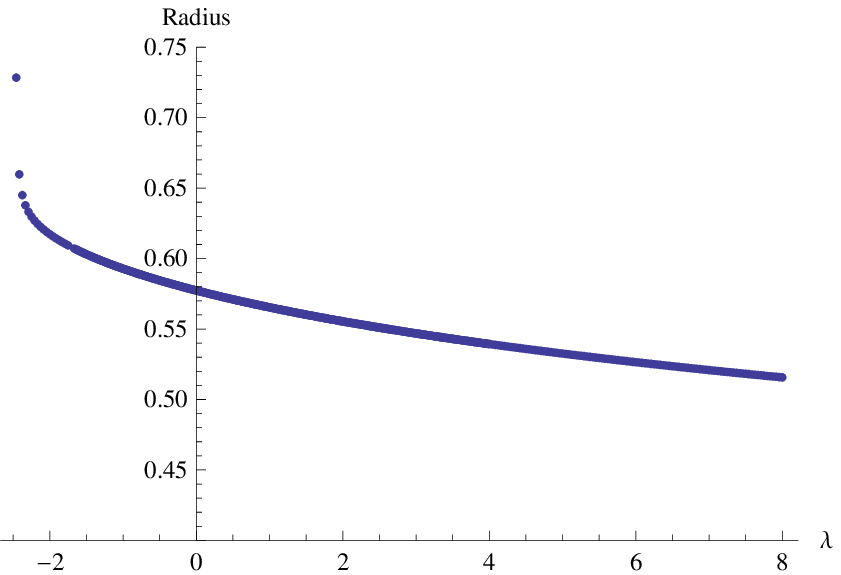,width=7cm}{Profiles of the Higgs
field for different $\lambda$, ranging from $\lambda v^2L^2 = 12.5$ on the top to $\lambda v^2 L^2
= -0.8$ on the bottom.\label{lambdaradius}}{Radius of the wall as function of
$\lambda$.\label{lambda}}
where $f(\beta)$ is defined by
\be f(\beta)= {\rm Min}_{\varphi} \int_1^{\infty} dx \left( x^4(\partial_x \varphi)^2 + \frac{\beta
x^2}{4}(\varphi^2-1)^2 \right) \ee
subject to the boundary conditions $\varphi(1)=0$ and $\varphi(\infty)=1$. We see that the
contribution from the scalar field and the vacuum energy always scale as $R^3$. The energy is
minimized at
\be R= \sqrt{\frac{BL^3}{ v}} \left({\frac{1}{3 f(\beta)+\beta/4}}\right)^{1/4} \label{rlam}\ee
To compute the function $f(\beta)$ we need one numerical integral subject to $f(0)=3$. The result
for the wall radius is shown in Figure \ref{lambda}  for fixed $B$. The fact that $R$ grows
(shrinks) with decreasing (increasing) $\lambda$ can be traced to the IR vacuum energy in
\eqn{lame}. The divergence of $R$ for $\lambda<0$ occurs numerically only when $\lambda  v^2 L^2 <
-9/4$, which is already beyond the BF bound and therefore outside the physically relevant regime.
For $\lambda
>0$, the wall position changes only gradually.

\para
For very large $\lambda$, one may neglect the contribution from the gradient energy and potential
energy for $r>R$ completely and the energy battle is dominated by the infra-red potential vs.
magnetic field. In this regime $\beta \gg f(\beta)$ in \eqn{rlam} and the wall position is given by
\be R \simeq \frac{ \sqrt{2 B}L}{\lambda^{1/4}v} \ \ \ \ \ \ \ \ \ \ \ \ \ \ \ \ \ \ \ \ \ \ \
((\lambda v^2L^2)^{1/4}  \gg 1)\label{rbig}\ee
However, there are two reasons why we should be cautious about this result. Firstly, at large
$\lambda \gg e^2/L^2\kappa^2v^4$, the gravitational backreaction becomes important and a black hole
forms. Secondly there is the possibility of a different monopole wall with $\phi\neq 0$ in the
infra-red \cite{private} which becomes a competitor for the wall discussed above at large
$\lambda$.

\subsubsection*{Further Properties}

Before we describe further properties of the wall in $AdS$, it will prove useful to first review
what (little) is known about the wall in flat space. When the potential vanishes, $\lambda=0$, the
monopole wall is BPS, obeying the first order Bogomolnyi equations
\be B_i^a = ({\cal D}_i\phi)^a\nn\ee
Suppose the wall sits at $z=0$. For $z\ll 0$, the solution is $B^a_i\approx 0$ and $\phi^a\approx
0$, up to exponentially small corrections. For $z\gg 0$, the wall emits a constant magnetic field
$B_z^3=B$. The Bogomolnyi equation requires that this magnetic field is supported by a linearly
growing scalar, $\phi^3 = Bz$. This means that in flat space there is no asymptotic expectation
value of the scalar field and the solution is characterized by just a single dimensional scale,
$B$.  In the neighbourhood of $z=0$, the solution is much more complicated and, in particular,
exhibits a lattice structure. Numerical evidence for this solution was presented in \cite{ward},
where the plots of energy density clearly reveal the lattice, with each fundamental domain carrying
one unit of monopole flux. We will discuss this lattice more in the next section. Here we would
first like to describe a few other features of the wall.

\para
Firstly, one may wonder about the relationship between our monopole wall in $AdS$ when $\lambda=0$
and the BPS wall in flat space. In contrast to flat space, the curvature of $AdS$ allows for the
scalar to asymptote to a constant expectation value \eqn{l0prof}. It appears therefore that our
wall solution depends on two dimensionful quantities: $v$ and $B$. However, this is a little
misleading since the radial position of the wall adjusts itself so that in the neighbourhood $r\sim
R$, the $AdS$ solution does indeed obey the Bogomolnyi equation. To see this we work in the
coordinates $r = R(1+z/L)$ where $z$ has been chosen to have a canonically normalized flat metric
in the vicinity of the wall. Expanding the $\phi$ profile \eqn{l0prof} gives
\be \phi \approx \frac{3v}{L}\, z\nn\ee
which indeed obeys the Bogomolnyi equation $\partial_z \phi = B_{\rm local}$ with the local flux
given in \eqn{lflux}. Note also that in flat space the monopole wall has linearly diverging energy
per unit of area. Embedding it into AdS regularizes the energy density to a finite value.

\para
Next we would like to estimate the thickness, $\delta$, of the wall which we will define by the
fall-off of the W-boson fields. Again, it is simplest to start in flat space. Here the $\phi$ gives
a mass to the W-bosons which grows linearly with distance from the wall. For $z$ suitably large,
the solution is given by $W \sim \exp(-\phi z) = \exp(-vz^2/L)$, where we have dropped numerical
factors in the exponent as we care only for the parameteric scaling. The width in flat space is
given by $\delta_{\rm local } \sim \sqrt{L/v}$. In $AdS$, the W-boson profile falls off as a
power-law, but with characteristic width again given, locally, by $\delta_{\rm local}$. Including
the $AdS$ warp-factor, the width of the domain wall in the radial direction is
\be \delta \sim \frac{R}{L}\,\,\delta_{\rm local} \sim \frac{L\sqrt{B}}{v}\nn\ee
With this thickness in hand, we should perform a consistency check that the flat space BPS monopole
wall can indeed be embedded in $AdS$. We require $\delta_{\rm local} \ll L$, so that the curvature
is negligible. This, in turn requires $\delta \ll R$ which can be guaranteed by taking
\be vL \gg 1\nn\ee
Note the the magnetic field $B$ disappears from this estimate: both $\delta$ and $R$ scales like
$\sqrt{B}$. If $vL \gg 1$ we are at liberty to take the flat space numerical solution for the
monopole wall studied in \cite{ward} and simply embed it in $AdS$; only the long distance tails
will be changed and these are well captured by the Abelian, bag approximation described above.

\section{The Physics of the Wall}

In this section we describe the physics of the boundary theory that results from the existence of
the wall. We start by describing the sources which arise from the leading order fall-off of the
various $U(1)$ gauge field and the scalar.

\subsection{Sources}

Since magnetic monopoles owe their existence to winding Higgs expectation value $v^a(x)$, one may
worry that this constitutes a source in the boundary theory that explicitly breaks translational
invariance. In fact, it does not:  the sources of the boundary theory are homogeneous. We now
explain this.

\para
Let us first briefly return to global $AdS$. In that case, we know that the expectation value of
$v^a$ must wind on the boundary. One can happily move the location of the winding through the use
of a large gauge transformation, retaining the condition ${\cal D}v^a=0$. However, one can only
remove the winding altogether, so that $v^a$ is strictly constant on the boundary, through a
singular gauge transformation which introduces Dirac strings into the Abelian gauge potential
$C_\mu$ defined in \eqn{longmag}. From the perspective of the boundary theory, $C_\mu$ is a source
and these Dirac strings appear as delta-function magnetic impurities.

\para
In contrast, in planar $AdS$ there is no obstacle to picking a gauge in which $v^a$ is strictly
constant on the boundary. In this case, the requisite winding all takes place behind the infra-red
horizon. This is simple to see using the construction above where we need only take the limit
\eqn{scaling} to zoom into a patch in which $v^a$ is locally constant.

\para
The upshot of this argument is that in any gauge ${\cal D}v^a=0$ on the planar boundary, but one
may choose a (non-singular) gauge in which $\partial v^a=0$.  The interpretation of $v^a$ for the
boundary theory is different for $\lambda=0$ and $\lambda\neq 0$. In the latter case, the winding
of the Higgs expectation value has no bearing on on the boundary theory: equation \eqn{dphi}
ensures that ${\cal D}_\mu v^a=0$ and this does not contribute the action \eqn{action} in any way
other than through the induction of a background magnetic field. We may then simply set the source
$\alpha$=0 in \eqn{phiexp}.

\para
The case of vanishing potential, $\lambda =0$, is a little more subtle since the vacuum expectation
$v^a(x)$ is now itself the source for the field theory. Clearly in the gauge $\partial_\mu
v^a(x)=0$, the source does not break translational invariance. However, the statement of
translational invariance should be a gauge invariant statement. In the Appendix we show that in
order to construct a conserved stress-energy tensor, we need only the weaker condition ${\cal
D}_\mu v^a(x)=0$, which is true in any gauge.

\para
The wall also gives rise to a source for the $U(1)$ current $J^\mu$ of the boundary theory.
(Strictly speaking, the logic of this sentence is backwards: we should say that the monopole wall
is a solution of the bulk equations in the presence of a source).  The long-range magnetic field
was defined in \eqn{longmag}. In the gauge in which $\partial_\mu v^a=0$ on the planar boundary,
this is given by $F_{\mu\nu} = \partial_\mu C_\nu -
\partial_\nu C_\mu$. The long range field arising from the wall is simply a constant $F_{12}=B$.
%
%
This has the simple interpretation of placing the boundary theory in a constant magnetic field
through the  coupling,
\be L_{\rm current} =  C_\mu J^{\mu}\label{abeliansc}\ee

\subsection{The Lattice Structure}

While the leading order fall-off of the various fields -- those that correspond to sources -- are
translationally invariant, the full solution is not. It forms a lattice-like structure. As we
explained previously, this is expected on general grounds from studies of monopoles with high
winding number and was confirmed by the numerical work of Ward \cite{ward}. The full profile of the
wall is not well understood, although recently the spectral curve for a monopole wall with square
lattice structure was obtained \cite{spectral}. Nonetheless, there are some basic lessons that we
can take away.

\para
The spatially modulated part of the solution is seen in the sub-leading terms in the asymptotic
expansion of fields. From the perspective of AdS/CFT, this means that the periodicity of the
monopole wall is imprinted on the expectation values: $\langle J_\mu\rangle$, dual to the bulk
photon; $\langle{\cal O}\rangle$, the operator dual to the bulk Abelian scalar; and  $\langle {\cal
W}_\mu(x) \rangle$, dual to the bulk W-boson. The  boundary theory {\it spontaneously} breaks
translational invariance, dynamically forming a lattice\footnote{We stress that this is spontaneous
breaking, rather than the explicit introduction of a lattice, either through D-branes \cite{shamit}
or sources \cite{dw,stripe}.}, analogous to a (baby) Skyrme crystal. Similar lattices are thought
to form in several condensed matter systems, including in the context of  quantum Hall
ferromagnetism  \cite{qhs} and graphene \cite{sarma}. Back in the holographic world, familiar
Abrikosov vortex lattices have been studied in superconductors in \cite{maeda}, while a spatially
modulated phase was found to arise from the instability of bulk electric fields in 5d Chern-Simons
theories \cite{ooguri1,ooguri2}.

\para
We do not currently know what form the lattice takes. The numerical studies \cite{ward} focus on a
square lattice but the lattice structure is, to a large extent, a moduli of the flat space
Bogomolnyi equations. In $AdS$, the presence of both the gravitational attraction and (where
relevant) a non-vanishing potential will remove these moduli and the resulting lattice structure
will be determined dynamically. As we mentioned previously, a similar situation arises for
Skyrmions where it is known that a hexagonal lattice structure is marginally preferable to a square
lattice. It remains an open problem to determine the structure of the lattice in the present case.

\para
Nonetheless, even without an explicit solution, we can estimate the crude form of the periodicity
and the amplitude of the spatial modulation. In the vicinity of the wall, there is only one scale
which can set the lattice spacing: in the local frame, it must be of order ${\Gamma_{\rm
local}}\sim 1/\sqrt{B_{\rm local}}$, with $B_{\rm local}$ given in \eqn{lflux}. Taking into account
the warp factor, the lattice spacing in the boundary theory is
\be \Gamma  = \frac{L}{R}\,{\Gamma_{\rm local}} \sim \frac{1}{\sqrt{B}}\nn\ee
which we could have anticipated simply because $B$ is the only relevant dimensionful quantity in
the boundary theory.

\para
To fully compute the expectation values which are sensitive to the spatial modulation is  beyond
our current ability. For now, we make do with a simple toy model. We focus on the Abelian scalar
$\phi$ with vanishing potential, $\lambda=0$. We assume that the wall imprints a spatial modulation
on $\phi$ at $r=R$, and solve the equation of motion subject to the boundary conditions
\be \phi(r\rightarrow\infty) = v\ \ \ \ ,\ \ \ \ \ \phi(x,y,r=R) = \sqrt{B_{\rm
local}}\sin\left(\frac{x}{\Gamma}\right)\sin\left(\frac{y}{\Gamma}\right)\label{wavy}\ee
where the amplitude of the oscillations at $r=R$ is again determined on dimensional grounds and we
have chosen a square lattice for simplicity. (There may of course be order one coefficients
multiplying $\phi$ which we neglect).  The solution to the scalar equation of motion in $AdS$ is
given by
\be \phi(x,y,r) = v\left(1-\frac{R^3}{r^3}\right) + \Theta\, \frac{
L^2}{r^{3/2}}I_{3/2}\left(\frac{\sqrt{2}L^2}{\Gamma
r}\right)\sin\left(\frac{x}{\Gamma}\right)\sin\left(\frac{y}{\Gamma}\right)\nn\ee
where $I_{3/2}$ is the modified Bessel function and the coefficient constant $\Theta$ is determined
by the boundary condition at the wall \eqn{wavy} to be $\Theta =
2B^{3/4}(3vL)^{1/4}\sqrt{\pi}e^{-\sqrt{6vL}}$. The $1/r^3$ fall-off determines the expectation
value $\langle {\cal O}\rangle$ of the operator dual to $\phi$,
\be \langle {\cal O}\rangle \sim
B^{3/2}\left[\frac{1}{(3vL)^{1/2}}-\frac{4}{3}(6vL)^{1/4}e^{-\sqrt{6vL}}
\sin\left(\frac{x}{\Gamma}\right)\sin\left(\frac{y}{\Gamma}\right)\right] \ee
The take-home lesson from this simple calculation is that the  spatial modulation is exponentially
suppressed by $vL$, the dimension of the W-boson operator ${\cal W}_\mu$. We expect that this
feature will continue to hold in a correct treatment for both $\langle {\cal O}\rangle$ and for
$\langle{\cal W}_\mu\rangle$ where the spatial modulation is the leading order contribution.

\subsection{Symmetry Breaking and the Infra-Red}

Translational invariance is not the only symmetry broken by the monopole wall. The W-boson fields
are sourced by magnetic monopoles and the expectation value $\langle{\cal W}_\mu(x)\rangle$ breaks
the global $U(1)$ symmetry. This is similar to the situation of holographic p-wave superconductors
\cite{pwave1,pwave2}. However, rather strangely, the symmetry breaking is not induced by a
background chemical potential, but instead by a background magnetic field. Usually one expects
magnetic fields to suppress superconductivity, not to induce it. In the bulk, the energy cost of
turning on the massive W-boson fields is compensated by the energy gained in the infra-red, where
the profile is simply $\phi \approx B \approx 0$. (It was suggested in \cite{erickim} that by
adjusting moduli one can also form a monopole wall solutions in which $\phi\approx 0$ on one side
of the wall, but with non-trivial non-Abelian gauge fields turned on. This would appear to have
higher energy in $AdS$).

\para
The fact that $\phi \approx 0$  tells us that the $SU(2)$ global symmetry is restored in the
infra-red. This is not uncommon in the context of holographic RG flows. Indeed, our potential with
$\lambda <0$ and no magnetic field similarly exhibits an $AdS$ domain wall, interpolating between
the broken phase in the UV and unbroken phase in the IR. More novel is the screening of magnetic
field, $B\approx 0$, in the infra-red. This seems to be a intimately tied to the existence of the
monopole wall.

\para
The restoration of the unbroken phase in the infra-red provides an opportunity for non-Abelian
gauge dynamics to play an important role. In $AdS$, the running of the gauge coupling is cut-off at
the scale $L$ and confinement occurs only occurs if the gauge coupling is strong at this point.
Expressed in terms of the effective strong coupling scale $\Lambda \sim v\exp(-1/e^2)$, confinement
occurs if $\Lambda L \gg 1$. In the present context, we should in addition require that $R\Lambda
\gg 1$, to ensure that the confinement phase has room to form behind the wall. It would be
interesting to understand the effect of this confining phase on the transport properties of the
bulk theory.
%

\subsection{Comparison to Holographic Superconductors}

It is instructive to compare the monopole wall with the familiar story of holographic
superconductors \cite{gubser,hhh,sean,gary}. Recall that a  field with charge $q$ and mass $m$ will
condense around the horizon of an extremal $AdS$ black hole if \cite{denef}
\be q^2\gamma^2 \geq 3 + 2m^2L^2\label{unstable}\ee
where $\gamma^2 = 2e^2L^2/\kappa^2$: $\kappa$ is the gravitational coupling; $e$ the gauge
coupling; $L$ the $AdS$ length scale.

\para
At first glance, it seems unlikely that the physics at play in the electric case will carry over to
magnetic solitons. The electrical instability arises (at least in part) because the effective mass
of the scalar field is reduced by the background gauge field $A_t$ near the black hole. Yet this
does not happen for the solitonic monopoles in the presence of a magnetic black hole. Moreover, a
lowering of the mass would simply result in the size of the soliton increasing and it is unclear
what it means for to have $m^2<0$ for solitons.

\para
Nonetheless, if we blindly ignore this intuition  we can road test the mass and charge of the
monopole against the criterion \eqn{unstable}. A single magnetic monopole has mass $M_m \sim
v/e^2$. To compare the charges, we should first perform an S-duality so that we replace the factor
of $q^2e^2$ in \eqn{unstable} with $1/e^2$.  Since we are only after a rough estimate, we ignore
the constant factor of 3. The criterion \eqn{unstable} suggests that the extremal black hole may be
unstable if
\be v^2\kappa^2 < e^2\nn\ee
In the next Section we will see that this is indeed the case. However, all evidence points to the
fact that the transition between the magnetic black hole and monopole wall is first order as
opposed to the second order transition seen in the electric case.

\section{Gravitational Backreaction and Phase Transitions}

In this section we describe the gravitating monopole wall. There is a long literature studying a
single gravitating monopole, both in flat space and $AdS$. The case of multi-monopoles is much more
difficult and has only been approached using the Abelian bag approximation \cite{bhbag}. Here we
also restrict to the Abelian approximation, now working asymptotically planar $AdS$. Moreover, for
the remainder of this Section we work with vanishing potential, $\lambda=0$.

\para
Our main interest is in understanding the competition between the monopole and the \RN black hole.
The latter is always a solution with the same asymptotics as the monopole wall (meaning that the
black hole and the wall have same sources in the boundary theory; the subleading terms which
dictate expectation values differ between the two). We will see that when backreaction is taken
into account, the monopole wall only exists for a range of the parameters. In Section 4.1, we focus
on zero temperature solutions and study how they evolve as we increase  $v^2\kappa^2/e^2$ which
governs the relative strength of the magnetic and gravitational forces. We will see that as $v$ is
increased, there is a critical value beyond which the monopole wall ceases to exist.

\para
In Section 4.2 we put the system at finite temperature and watch the monopole wall become engulfed
by the black hole as we turn up the heat. From the perspective of the boundary theory, this is the
lattice melting transition.

\subsection{Ground State at $T=0$}

We start by describing the ground state dynamics at zero temperature. The gravitating monopole wall
is again a solution involving the Abelian fields, patched at $r=R$. We first discuss the
gravitating wall in the ultra-violet region, $r>R$, where the metric takes the general form,
\be ds^2 = -C(r)dt^2 + \frac{dr^2}{D(r)} + \frac{r^2}{L^2}(dx^2 + dy^2)\ \ \ \ \ \ \ \
r>R\label{uvmet}\ee
The Maxwell equations in this background allow for a constant magnetic field $B$ while, in the
absence of a scalar potential, the field $\phi$ satisfies
\be \partial_r (r^2 \sqrt{C(r)D(r)}\,\partial_r\phi)=0\label{scalar}\ee
The metric functions are determined by Einstein's equations. Suitable combinations of the $tt$,
$\theta\theta$, and $rr$ components yield the first order differential equations,
\be \partial_r\left(\frac{C}{D}\right)&=&\frac{\kappa^2}{e^2}\left(\frac{C}{D}\right)
\,r(\partial_r\phi)^2\label{e1}\\
\frac{\partial_rD}{r}+\frac{D}{r^2}-\frac{3}{L^2}&=&-\frac{\kappa^2}{2e^2}\,
\left(\frac{B^2L^4}{r^4}+D(\partial_r\phi)^2\right)\label{e2}\ee
It can be shown that, together with the matter equations, these imply the remaining, more
complicated, second order differential equation.
%
%

\para
We first integrate the equation of motion \eqn{scalar} for the scalar field, to get
\be \partial_r\phi = \frac{\alpha}{r^2\sqrt{CD}}\label{aphid}\ee
The integration constant $\alpha$ feeds into the the equations of motion for the metric
coefficients. We may solve these equations in an expansion around large $r$, which we write as
\be C(r) = \frac{r^2}{L^2} - \frac{2m}{r}+\ldots \ \ \ ,\ \ \ D(r)=\frac{r^2}{L^2} -
\frac{2m}{r}+\ldots\nn\ee
The leading term means that the space is asymptotically $AdS_4$ while the equations of motion
ensure that the next term -- characterized by a constant of integration, $m$ -- is the same for
both $C$ and $D$. The two functions differ in their subleading terms, denoted $\ldots$ above, which
depend on both $m$ and $\alpha$.

\para
The next step is to fix the integration constants, $m$ and $\alpha$, in terms of the microscopic
parameters $B$ and $v$. To do this, we first need to understand how (and where) the metric is
patched at $r=R$.

\para
In the infra-red, $r<R$, the field profiles are  $\phi=B=0$ and the metric is simply a slice of
$AdS$. To patch this onto \eqn{uvmet}m one must allow for a rescaling of the time coordinate. Such
rescaling is expected when one finds an emergent $AdS$ regime and corresponds to a lower effective
speed of light, $c_{\rm eff}$, in the infra-red \cite{rocha}. The metric takes the form,
\be ds^2 = \frac{r^2}{L^2}(-c_{\rm eff}^2\,dt^2 + dx^2 + dy^2) + \frac{L^2}{r^2}dr^2\ \ \ \ \ \ \
r<R\nn\ee
This is patched onto the metric \eqn{uvmet} by the standard Israel junction conditions which,
firstly, require that all metric components are continuous. The continuity of the $g_{rr}$
component provides a definition of the wall position, $R$, namely
\be D(R) = \frac{R^2}{L^2}\label{DR}\ee
This allows us to implicitly determine $R(m,\alpha)$. Meanwhile, continuity of the $g_{tt}$
component fixes the effective speed of light in the infra-red,
\be c_{\rm eff}^2 = \frac{C(R)}{D(R)}\nn\ee
The first of the Einstein equations \eqn{e1} tells us that $C(r)/D(r)$ is a monotonically
increasing function, ensuring that $c_{\rm eff} <1$.

\para
It still remains to fix the two integration constants, $m$ and $\alpha$. One of these is determined
by the boundary conditions for $\phi(r)$, namely $\phi(R)=0$ and $\phi(\infty)=v$. The remaining
integration constant is again fixed by the Israel junction conditions. In general, the extrinsic
curvature on either side of a domain wall must jump by an amount proportional to the tension of the
domain wall. However, in our case, the monopole wall is tensionless: the energy density increases
by a step-function at $r=R$, but there is no delta-function contribution. This ensures that the
extrinsic curvature is continuous which, in our parameterization, means that $\partial_rC(r)$ is
continuous at $r=R$. In contrast, $\partial_r D(r)$ will not be continuous on the wall.

\FIGURE{  \epsfig{file=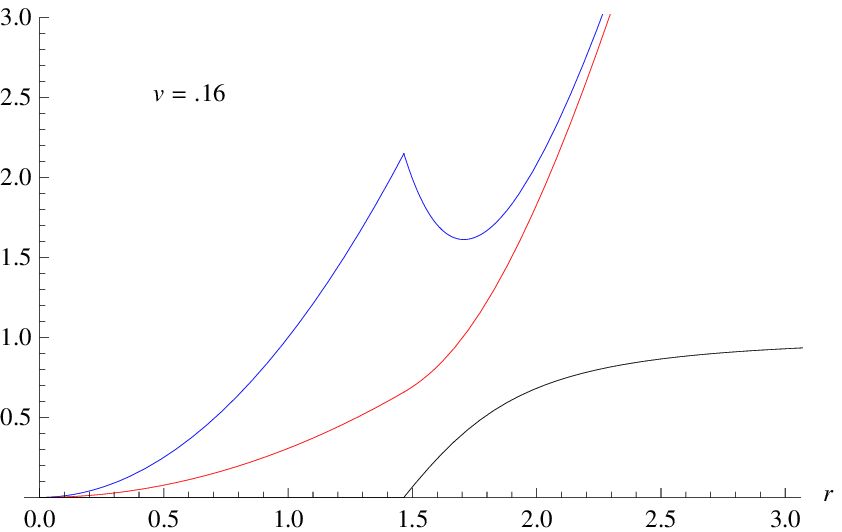,width=4.5cm} \qquad \epsfig{file=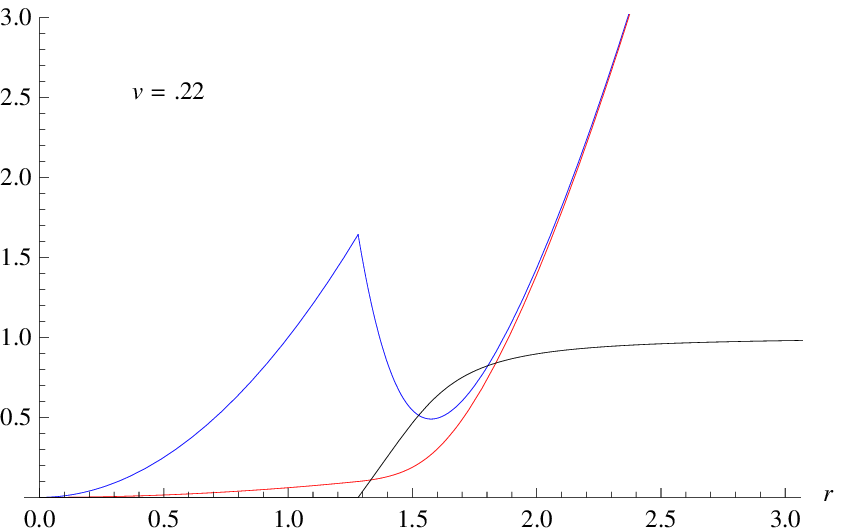,width=4.5cm} \qquad
\epsfig{file=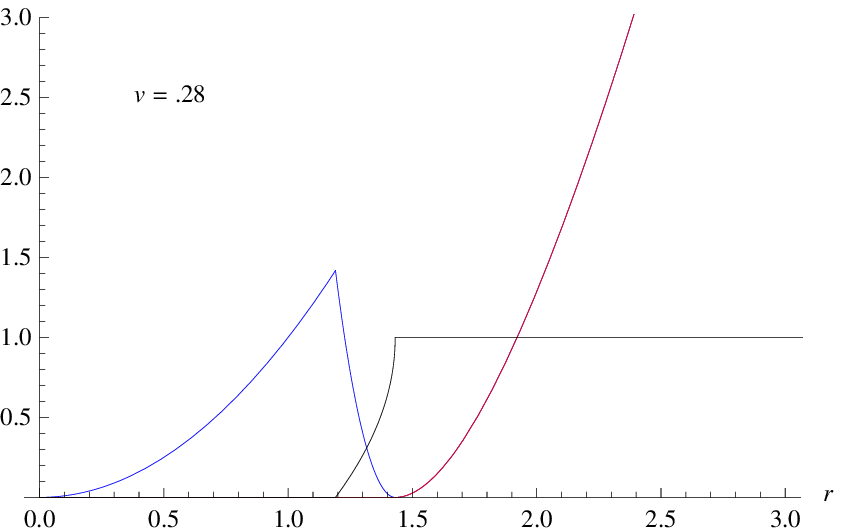,width=4.5cm}\caption[]{Profiles for $D(r)$ (blue), $C(r)$ (red) and
$\phi(r)$ (black) for increasing values of $v$. These plots have $L=B=e=1$ and
$\kappa^2=8\pi$}\label{zerotemperature}}
\para
The strategy outlined above is easy to implement numerically. Note that the equations are inariant
under the rescaling
\be r \to L\sqrt{\frac{B k}{e}} r \qquad \phi \to \frac{k}{e} \phi \ee
We can thus set the four parameters $L,e,B$ to one and $k^2=8\pi$ (this correspond to $G_N =1$). We
are thus presenting the plots for this choice of parameters. Different choices are related by a
simple rescaling. A typical plot of the profile functions is shown in Figure
\ref{zerotemperature}a. From top to bottom, the curves show $D(r)$, $C(r)$ and $\phi(r)$. The
position of the wall is indicated by the kink in $D(r)$ which coincides with $\phi(r)=0$.

\subsubsection*{Zero Temperature Phase Transition to the Black Hole}

As $v$ increases, the position of the wall shrinks into the infra-red. Eventually, the back
reaction on the geometry is enough to form a black hole. Our whole set-up remains, for now, at zero
temperature and the resulting black hole is extremal.

\para
Let us first come up with a rough and ready estimate for where this transition will take place.
Recall that in the absence of backreaction, the position of the wall  is given by \eqn{r0}
\be R^4 \sim \frac{B^2L^6}{9v^2}\nn\ee
We can compare this to the horizon of the extremal, magnetically charged, \RN black hole which sits
at
\be R_\star^4 = \frac{B^2L^6\kappa^2}{6e^2}\label{rstar}\ee
Notice that both positions scale in the same way with $B$; cranking up the magnetic field on the
boundary will not form a black hole in the bulk. In contrast, when $v$ gets large enough a black
hole does form. Naively, one might imagine that this happens when $R= R_\star$. We will see shortly
that this isn't quite true, although parameterically the black hole does indeed occur when
\be \frac{v^2}{e^2} > \frac{1}{\kappa^2}\nn\ee
It is worth noting that the weak gravity conjecture \cite{weak} explicitly rules out this regime of
parameters in a consistent theory of gravity. Indeed, one of the motivations for the conjecture was
the requirement that the minimally charged magnetic monopole is not an extremal black hole.
However, because $B$ plays no role in the transition to the black hole, this immediately translates
into the requirement that no horizon forms around the wall either.

\DOUBLEFIGURE[]{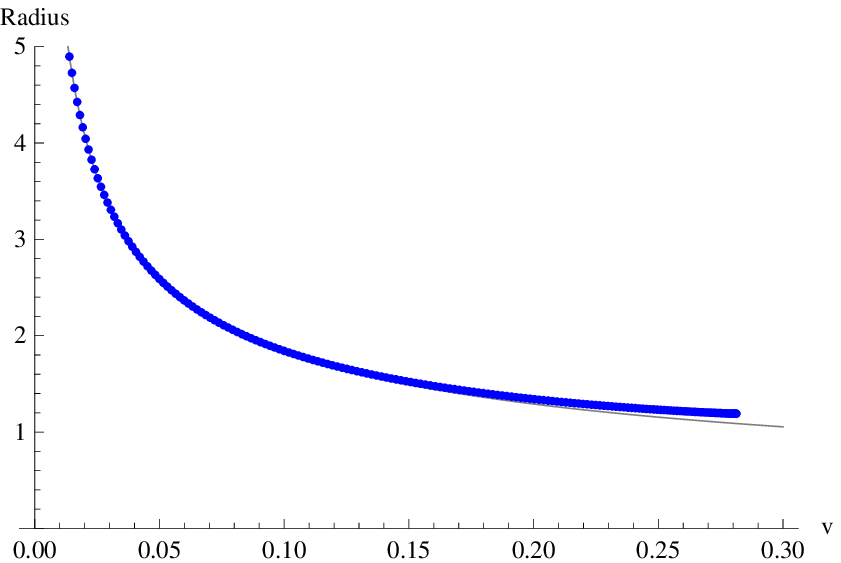,width=6.5cm}{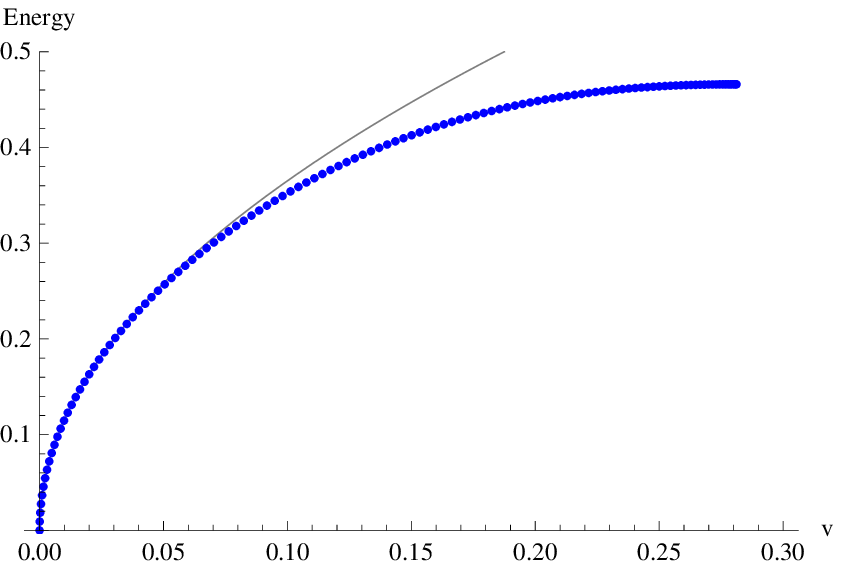,width=6.5cm}{Radius of the monopole
wall as function of $v$. We also plot the value $R=1/\sqrt{3 v}$ without
backreaction.\label{radiuszerot}}{The energy at zero temperature as function of $v$. At $v=v_{cr}$
becomes equal to the extremal black hole.\label{actionzerot}}
\para
The weak gravity conjecture notwithstanding, we can happily study the transition from the wall to
the black hole in the context of classical gravity. Figure \ref{zerotemperature} shows the profiles
function of the monopole wall as $v$ is increased\footnote{Full numerical disclosure: When
constructing these plots it is convenient to deviate somewhat from the strategy outlined in the
text. Instead of fixing $v$, we instead fix $\alpha$ and $R$. We  integrate \eqn{aphid} from $r=R$
to the boundary to find $v$ and then pretend that this was the $v$ we wanted all along. We use a
similar trick for $c_{\rm eff}$.}. The three figures have $v^2\kappa^2/e^2\approx 0.64$, $1.21$ and
$1.97$. The last plot in the figure is the critical value of $v$ at which a zero of $C(r)$ and
$D(r)$ appears at $r=R_\star$, signalling the formation of an extremal black hole. Notice that the
black hole horizon $R_\star$ does not form at the monopole wall, but instead at $R_\star>R$. From
this point onwards, the metric functions outside the horizon are given by the familiar extremal
black hole
\be C(r)=D(r) = \frac{r^2}{L^2}\left(1-4\left(\frac{R_\star}{r}\right)^3 +
3\left(\frac{R_\star}{r}\right)^4\right)\ \ \  \ \ \ \ \ \ r>R_\star\label{rnbh}\ee
Notice that as we approach the critical value of $v^2\kappa^2/e^2 \approx 1.97$, the profile for
the scalar field becomes flatter and flatter. When the black hole forms, the scalar field is
constant outside the horizon.

\para
In Figure \ref{radiuszerot}  we plot the position of the wall as a function of $v$.  We see that
the position of the gravitating wall agrees well with the estimate \eqn{r0} ignoring backreaction.
Notice that at the critical point ($v\approx 0.28$ in the figure) the radius of the wall is indeed
inside the horizon $R_\star = \left({\frac{B^2L^6\kappa^2}{6e^2}}\right)^{1/4}$.


\para
In Figure \ref{actionzerot} we plot the energy of the black hole solution. (One must subtract the
energy of a reference background which is pure $AdS$). The energy is strictly less than that of an
extremal \RN black hole, but becomes equal at the critical value $v^2\kappa^2/e^2 \approx 1.97$. We
stress that the extremal \RN black hole \eqn{rnbh} is a solution for all values of $v$, but is
energetically disfavoured below the critical value. However, unlike the story of electrically
charged \RN AdS solutions, the black hole does not appear to have a linear instability. (At least,
not one that we are aware of).


\subsection{Turning up the Heat}

It is a simple matter to heat up the boundary theory. The infra-red $AdS$ region is replaced by
Schwarzchild $AdS$, again with $B=\phi=0$.
\be ds^2 = -C(r) dt^2 + \frac{dr^2}{D(r)}+ \frac{r^2}{L^2}(dx^2 + dy^2) \ \ \ \ \ \ \ \ \ \ \ \
r<R\nn\ee
with
\be D(r)= \frac{C(r)}{c_{\rm eff}^2} = \frac{r^2}{L^2}\left(1-\left(\frac{R_{\rm hor}}{r}\right)^3
\right)\label{irbh}\ee
%
%
%
This is patched onto the monopole wall solution at $r=R$, where the position $R$ of the wall will
be determined dynamically by the junction conditions which are once again $D(R)$, $C(R)$ and
$C^\prime(R)$ continuous. We will study the transition between this monopole wall and the \RN black
hole as $v$ is increased.

\para
A quick comment: the full non-Abelian solution is again expected to form a crystal, breaking both
translational invariance and the global $U(1)$ symmetry of the boundary theory. As usual in
AdS/CFT, the large $N$ nature of the boundary field theory protects this long range order from the
expected quantum fluctuations in two dimensions.

\para
The presence of the red-shift factor $c_{\rm eff}$ in the metric \eqn{irbh} means that it also
appears in the temperature, $T$, of the boundary theory which is given by
\be T = \frac{3c_{\rm eff}(v) R_{\rm hor}}{4\pi L^2}\label{temp}\ee
\FIGURE{\epsfig{file=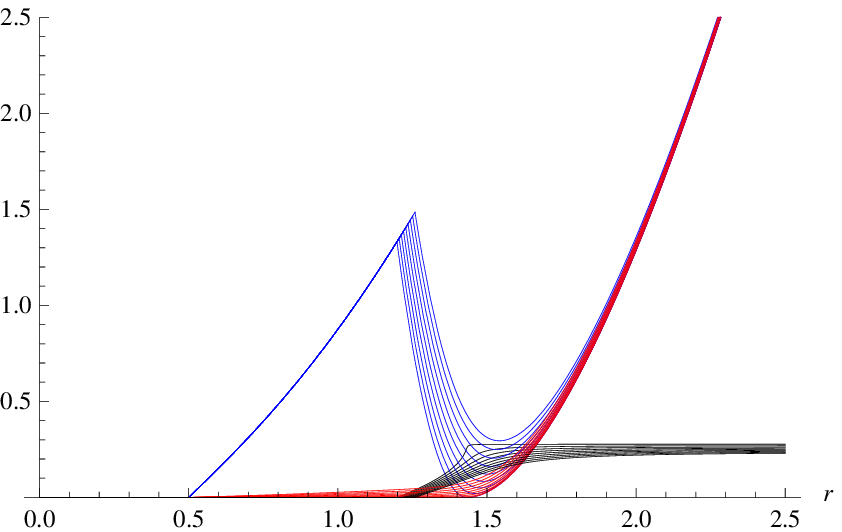,width=4.5cm} \epsfig{file=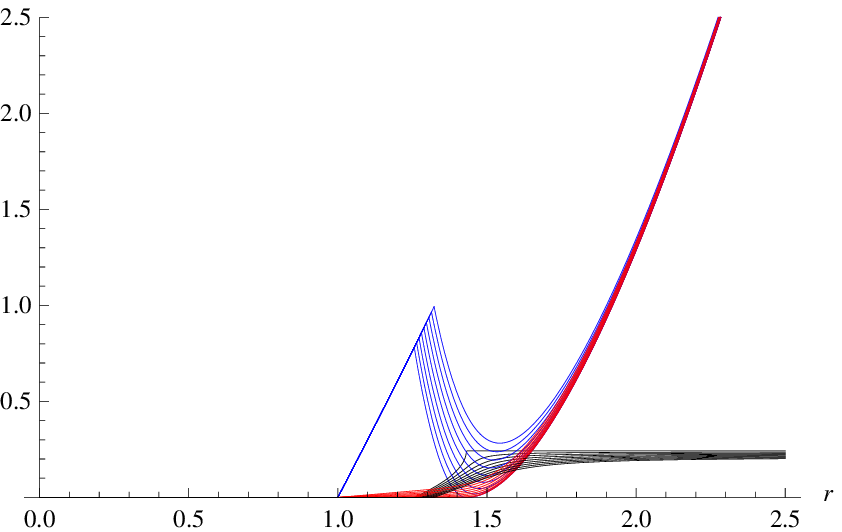,width=4.5cm}
\epsfig{file=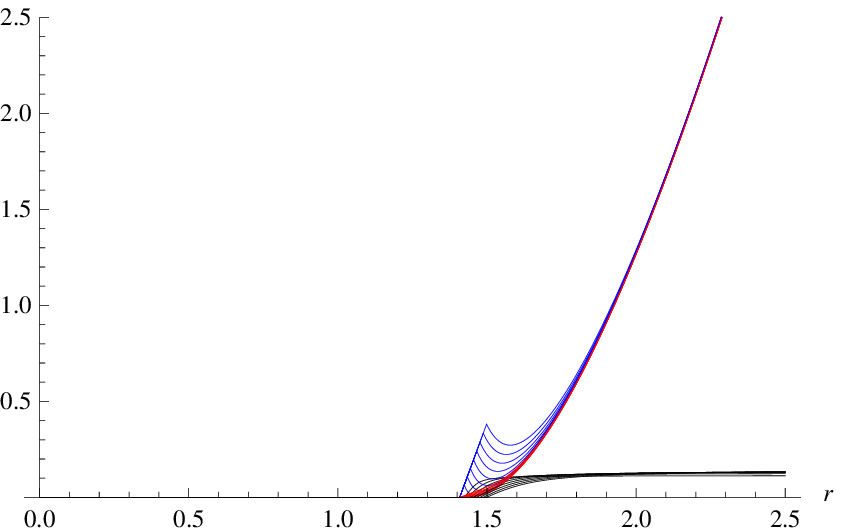,width=4.5cm}
        \caption[]{Monopole-black hole transition for increasing values of $R_{\rm hor}$.}%
\label{plott}}
This causes a minor complication when solving the equations of motion numerically. The strategy we
employ is the simplest one: we fix $R_{\rm hor}$ and study the solutions as we vary $v$. However,
$c_{\rm eff}$ depends on $v$ which means that we are not looking at isothermal deformations. Below
we will first present the numerical results $R_{\rm hor}$-$v$ plane and subsequently describe the
qualitative features in the more physical $T$-$v$ plane.

\para
Let us summarize the different radial positions in the game: $R_{\rm hor}$ is the IR Schwarzshild
horizon; $R$ is the position of the monopole wall; $R_\star$ is the position of the extremal \RN
black hole given in \eqn{rstar}.

\subsubsection*{Small $R_{\rm hor}$}

In what follows, we fix $B$, $L$ and $\kappa$. This fixes $R_\star$ to be given by \eqn{rstar} and
we begin our discussion by looking at the regime $R_{\rm hor} < R_\star$. The transition from the
monopole wall to the black hole (as we crank up $v$) happens in much the same way as in the
previous section: when $v$ reaches a critical value a horizon forms outside the monopole wall at
$R_\star$. The critical value of $v$ decreases as $R_{\rm hor}$ increases and the transition occurs
when the monopole wall sits in the window $R_{\rm hor} < R(v) < R_\star$. The resulting \RN black
hole is {\it always extremal}. This illustrates the point we made above: changing $v$ is not an
isothermal deformation. In the present case, the transition always occurs at zero temperature, even
if $R_{\rm hor}\neq 0$. The vanishing temperature arises due to the presence of $c_{\rm eff}(v)$ in
\eqn{temp} which tends towards zero as $v$ tends towards its critical value. Figure \ref{plott}
shows some typical examples of the transition with progressively bigger $R_{\rm hor}$. In the last
figure, $R_{\rm hor}$ is almost, but not quite, equal to  $R_*$.

\FIGURE{\epsfig{file=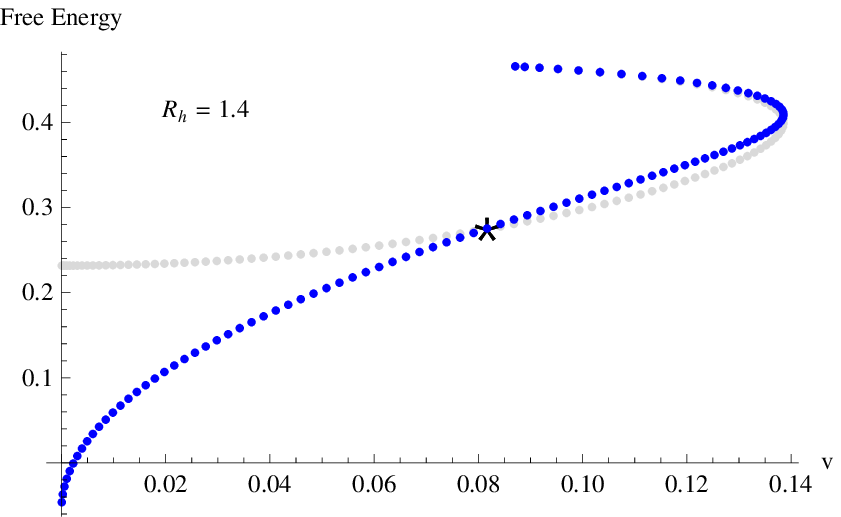,width=6.5cm} \caption[]{The free energy as function of $v$ for
$R_{\rm hor}$ less than, but close to, $R_*$.\label{actiont}}}
\para
The free energy of the solution is plotted (in blue) in Figure \ref{actiont} as a function of $v$.
For small $v$, the free energy is negative. This arises from the  IR Schwarzchild black hole which
has $F=-R_{\rm hor}^3/2\kappa^2L^4$. As $v$ increases, the contribution from the monopole wall
becomes more important. It can be checked that the free-energy at the end point of the curve agrees
with that of an extremal \RN black hole.

\para
The most surprising feature of the plot is that the free-energy is double-valued. This reflects the
fact that, within a certain parameter range, there are two solutions with the same asymptotics ($B$
and $v$) but different $R_{\rm hor}$. To understand the physics, we should look at the grey line
Figure \ref{actiont}; this shows the free energy of the non-extremal \RN black hole which sits at
the same temperature as the monopole wall, namely $T\sim c_{\rm eff} R_{\rm hor}$. The phase
transition between the wall and the black hole occurs where the two curves intersect. The
transition from the wall to monopole is always first order and  always occurs before we reach
before the turning point of the curve. A natural interpretation for the upper part of the curve
(where it is double valued) is that it corresponds to the (linearly) unstable solution.

\subsubsection*{Large $R_{\rm hor}$}

\FIGURE{ \epsfig{file=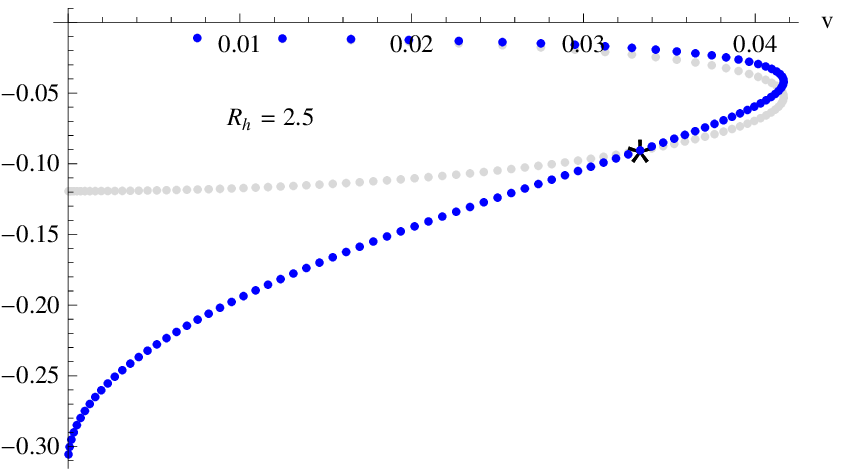,width=6.5cm} \caption[]{The free energy as function of $v$ for
different $R_{\rm hor}>R_*$.}
	\label{actiontt}}
For $R_{\rm hor}> R_\star$, the transition occurs in a different manner: as we increase $v$, the
position of the monopole wall decreases until it hits the horizon: $R=R_{\rm hor}$. But this is now
the horizon of a non-extremal \RN black hole, with temperature
\be T= \frac{R_{\rm hor}}{4\pi L^2}\left(3-\frac{L^2B^2\kappa^2}{2R_{\rm hor}^4e^2}\right)\nn\ee
A typical plot of the free energy is shown in Figure \ref{actiontt}. At the end point, the free
energy coincides with that of the non-extremal black hole. However, if one attempts to follow the
blue line in a thermodynamic process, we never reach the end point, nor even the turning point. The
transition to the black hole with temperature $T\sim c_{\rm eff}(v) R_{\rm hor}$ once again happens
at the crossing of the blue and grey lines.
%

\subsubsection*{The $T$-$v$ Phase Diagram}

\FIGURE{\epsfig{file=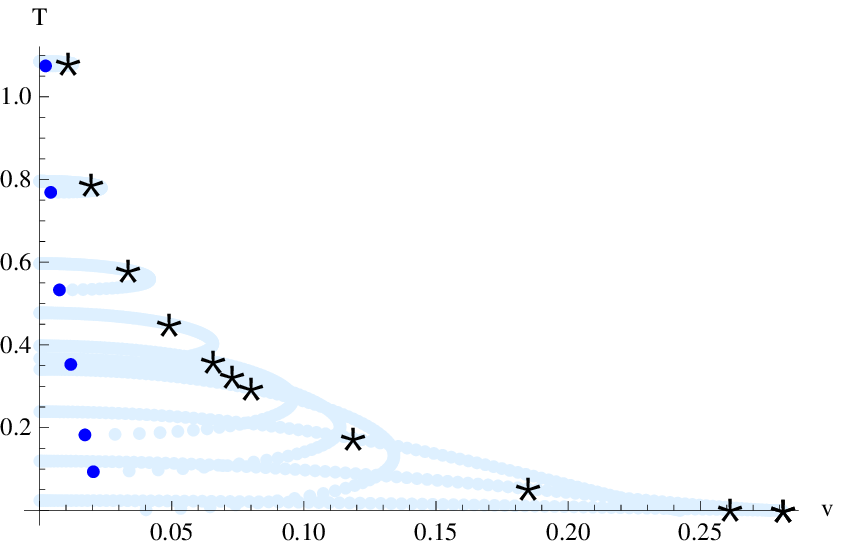,width=6.0cm}
        \caption[]{The phase diagram in the $T$-$v$ plane.}%
	\label{tvfig}}
From the previous analysis we can infer the gross features of the phase diagram in the $T$-$v$
plane. The \RN black hole exists everywhere in the phase diagram; the monopole wall exists only in
the lower left-hand corner of Figure \ref{tvfig}. The light-blue curves are reconstructions of the
monopole free-energy, similar to those plotted (in dark blue) in Figures \ref{actiont} and
\ref{actiontt}. The crosses again represent where the monopole wall and black hole have equal
energy; the monopole wall is thermodynamically stable to the left of the line of crosses.

\para
There are two other lines of significance shown in the Figure. The line formed by joining together
the dark blue dots shows where the second, unstable, monopole wall begins to exist. The envelope
formed by the light-blue curves shows where the monopole wall solution ceases to exist: the first
and second branch of monopole solutions annihilate here.  To the right of this there is only a
black hole. Notice that the crosses coincide with this envelope at zero temperature. This suggests
that the first order phase transition becomes second order as $T\rightarrow 0$.


\section*{Appendix: Symmetries and Sources}
\setcounter{section}{1} \setcounter{equation}{0}
\renewcommand{\theequation}{\Alph{section}.\arabic{equation}}

The purpose of this appendix is to discuss the somewhat more subtle case of $\lambda=0$, in which
the expectation value appears as a source for the field theory operators,
\be L_{\rm scalar} = v^a(x)\Phi^a(x) \label{scalarsc}\ee
Naively, it looks as if any winding $\partial_iv^a\neq 0$ will break translational invariance. Our
goal here is to show that this isn't the case and the weaker condition ${\cal D}_iv^a$ is enough to
ensure that there still exists a conserved stress tensor arising from translational invariance.

\para
Let's first discuss the fate of the global $SU(2)$ symmetry. $\Phi^a$ transforms in a triplet of
$SU(2)$ which means that, the presence of the source \eqn{scalarsc}, the current obeys
\be \partial_\mu J^{a\mu} = \epsilon^{abc}v^b\Phi^c\nn\ee
If the source is spatially uniform then a residual $U(1)$ symmetry survives, with the corresponding
current given by $\tilde{J}_\mu = v^aJ_\mu^a$. If, however, the source varies over space, then it
appears that the $SU(2)$ symmetry is now fully broken, with the divergence of $\tilde{J}_\mu$ given
by
\be \partial_\mu \tilde{J}^\mu = (\partial_\mu v^a)J^{a\mu}\ee
However, the symmetry is rescued by a background source for the current
\be L_{\rm current} =  A_\mu^a(x)J^{a\mu}\label{currentsc}\ee
As usual, it may be that the coupling to the current is more complicated than this, as happens for
a free scalar field. In all cases the coupling above should be viewed schematically. In general the
requirement is that a gauge transformation of the background field $A_\mu^a$ can be compensated by
a suitable redefinition of the operators in the theory. The $SU(2)$ currents now obey
\be {\cal D}_\mu J^{a\mu} \equiv \partial_\mu J^{a\mu} + \epsilon^{abc} A_\mu^bJ^{c\mu} =
\epsilon^{abc}v^b\Phi^c\nn\ee
We see that by playing the two sources $v$ and $A_\mu$ off against one another we can save a global
$U(1)$ symmetry even in the presence of a winding source. We have
\be
\partial_\mu \tilde{J}^\mu = ({\cal D}_\mu v^a) J^{a\mu} \nn\ee
which  is conserved provided that ${\cal D}_\mu v^a\equiv \partial_\mu v^a + \epsilon^{abc}A_\mu^b
v^c=0$

%

\subsubsection*{Translational Symmetry}

We now turn to the existence of translational invariance of the boundary theory in the case of
planar $AdS$. (A similar analysis holds for rotational invariance in global $AdS$. The trick is to
compensate a translation with a suitable gauge transformation so that one can define a conserved
momentum. We start by imagining that the background sources, $v^a$ and $A_\mu$ are dynamical so
that they too transform under translations and $SU(2)$ global rotations,
\be \delta v^a &=& \eta^\nu (\partial_\nu v^a + \epsilon^{abc}\alpha_\nu^b v^c) \nn\\
\delta A^a_\mu &=& \eta^\nu(\partial_\nu A^a_\mu - {\cal D}_\mu\alpha_\nu^a)\nn\ee
Here we take $\eta$ to be infinitesimal, with $\alpha_\nu$ parameterizing the $su(2)$
transformation.  For such a theory, the Lagrangian transforms in the usual fashion as $\delta {\cal
L} = \eta^\nu \partial_\nu {\cal L}$ and we can define a conserved energy-momentum tensor
$T^{\mu\nu}$ obeying $\partial_\mu T^{\mu\nu}=0$.

\para
However, in our world $v$ and $A_\mu$. They are fixed background sources. As such, the change of
the Lagrangian picks up extra terms,
\be \delta{\cal L} &=& \eta^\nu\left( \partial_\nu{\cal L} - \frac{\partial L}{\partial
A_\mu^a}\,\delta_\nu A^a_\mu-\frac{\partial {\cal L}}{\partial v^a}\,\delta_\nu v^a \right) \nn\\
&=& \eta^\nu\left(\partial_\nu {\cal L} - J^{a\mu}(\partial_\nu A_\mu^a + {\cal D}_\mu\alpha_\nu^a)
- \Phi^a (\partial_\nu v^a + \epsilon^{abc}\alpha_\nu^b v^c)\right) \nn\ee
This simplifies through a cunning choice of gauge transformation. We write
\be \alpha_\nu^a = A_\nu^a + v^a \beta_\nu\nn\ee
where the functions $\beta_\nu$ are, for now, left undetermined. Using the fact that ${\cal D}_\nu
v^a=0$, we find that the stress-energy tensor obeys
\be \partial_\mu T^{\mu\nu} = J_\mu^aF^{a\mu\nu} + v^aJ^a_\mu\partial^\mu\beta\nn\ee
There is one final fact that we need about the gauge field. As explained in Section 2, the
condition ${\cal D}_\mu v^a=0$ ensures that the magnetic field sits entirely in the unbroken
$U(1)$: $F^a_{\mu\nu} = -\hat{v}^a F_{\mu\nu}$ where $F_{\mu\nu}$ was given in \eqn{longmag}. Using
this, we get our final expression for the divergence of the stress tensor
\be \partial_\mu T^{\mu\nu} = \tilde{J}_\mu (-F^{\mu\nu} + \partial^\mu\beta^\nu)\nn\ee
We have a conserved momentum only if we can find functions $\beta_\nu$ obeying $\partial_\mu
\beta_\mu = G_{\mu\nu}$. This set of equations is overly constrained and, in general, does not have
solutions. The exception is when $F_{\mu\nu}$ is constant on the boundary. For example, in the
context of the monopole wall, we have $F_{12}= B$ and the simple solution
$\beta_i=\epsilon_{ij}x^jB$

\section*{Acknowledgement}
Our thanks to Ofer Aharony, Micha Berkooz, Nick Dorey, Andrew Green, Sean Hartnoll, Andreas Karch,
John McGreevy, Harvey Reall, Simon Ross, Richard Ward and Shimon Yankielowicz for useful
discussions on this and related projects. We are grateful to Paul Sutcliffe for permission to
reproduce Figures \ref{stolen} and \ref{stolen2}. DT is supported by the Royal Society.


\begin{thebibliography}{99}

\small
\parskip=0pt plus 2pt

\bibitem{hongmag}   N.~Iqbal, H.~Liu, M.~Mezei and Q.~Si,
  ``{\it Quantum phase transitions in holographic models of magnetism and
  superconductors},''
Phys.\ Rev.\  D {\bf 82}, 045002 (2010) [arXiv:1003.0010 [hep-th]].
\bibitem{bag}  S.~Bolognesi,
  ``{\it Multi-monopoles and magnetic bags},''
  Nucl.\ Phys.\  B {\bf 752}, 93 (2006)
  [arXiv:hep-th/0512133].
\bibitem{ward}   R.~S.~Ward,
  ``{\it A monopole wall},''
  Phys.\ Rev.\  D {\bf 75}, 021701 (2007)
  [arXiv:hep-th/0612047].
\bibitem{gubser}  S.~S.~Gubser,
  ``{\it Breaking an Abelian gauge symmetry near a black hole horizon},''
  Phys.\ Rev.\  D {\bf 78}, 065034 (2008)
  [arXiv:0801.2977 [hep-th]].
\bibitem{hhh}  S.~A.~Hartnoll, C.~P.~Herzog and G.~T.~Horowitz,
  ``{\it Building a Holographic Superconductor},''
  Phys.\ Rev.\ Lett.\  {\bf 101}, 031601 (2008)
  [arXiv:0803.3295 [hep-th]]; ``{\it Holographic Superconductors},''
  JHEP {\bf 0812}, 015 (2008)
  [arXiv:0810.1563 [hep-th]].
\bibitem{sean}   S.~A.~Hartnoll,
  ``{\it Lectures on holographic methods for condensed matter physics},''
  Class.\ Quant.\ Grav.\  {\bf 26}, 224002 (2009)
  [arXiv:0903.3246 [hep-th]].
\bibitem{gary}  G.~T.~Horowitz,
  ``{\it Introduction to Holographic Superconductors},''
  arXiv:1002.1722 [hep-th].
\bibitem{Hartnoll:2007ai}
  S.~A.~Hartnoll and P.~Kovtun,
  ``{\it Hall conductivity from dyonic black holes,}''
  Phys.\ Rev.\  D {\bf 76} (2007) 066001
  [arXiv:0704.1160 [hep-th]].
\bibitem{sachdev}   S.~Sachdev,
  ``{\it Holographic metals and the fractionalized Fermi liquid},''
  Phys.\ Rev.\ Lett.\  {\bf 105}, 151602 (2010)
  [arXiv:1006.3794 [hep-th]].
\bibitem{pwave1}   S.~S.~Gubser,
  ``{\it Colorful horizons with charge in anti-de Sitter space},''
  Phys.\ Rev.\ Lett.\  {\bf 101}, 191601 (2008)
  [arXiv:0803.3483 [hep-th]].
\bibitem{pwave2} S.~S.~Gubser and S.~S.~Pufu,
  ``{\it The gravity dual of a p-wave superconductor},''
  JHEP {\bf 0811}, 033 (2008)
  [arXiv:0805.2960 [hep-th]].
\bibitem{harvey}   J.~A.~Harvey,
  ``{\it Magnetic monopoles, duality, and supersymmetry},''
  arXiv:hep-th/9603086.
\bibitem{schap}  A.~R.~Lugo and F.~A.~Schaposnik,
  ``{\it Monopole and dyon solutions in AdS space},''
  Phys.\ Lett.\  B {\bf 467}, 43 (1999)
  [arXiv:hep-th/9909226];
\bibitem{schap2}
  A.~R.~Lugo, E.~F.~Moreno and F.~A.~Schaposnik,
  ``{\it Monopole solutions in AdS space},''
  Phys.\ Lett.\  B {\bf 473}, 35 (2000)
  [arXiv:hep-th/9911209].
\bibitem{Allahbakhshi:2010ii}
  D.~Allahbakhshi and F.~Ardalan,
  ``{\it Holographic Phase Transition to Topological Dyons,}''
  arXiv:1007.4451 [hep-th].
\bibitem{Radu:2004ys}
  E.~Radu and D.~H.~Tchrakian,
  ``{\it New axially symmetric Yang-Mills-Higgs solutions with negative
  cosmological constant},''
  Phys.\ Rev.\  D {\bf 71} (2005) 064002
  [arXiv:hep-th/0411084].

\bibitem{kapustin}   V.~Borokhov, A.~Kapustin and X.~k.~Wu,
  ``{\it Topological disorder operators in three-dimensional conformal field
  theory},''
  JHEP {\bf 0211}, 049 (2002)
  [arXiv:hep-th/0206054].
\bibitem{witten}   E.~Witten,
  ``{\it SL(2,Z) action on three-dimensional conformal field theories with Abelian
  symmetry},''
  arXiv:hep-th/0307041.
\bibitem{ross}   D.~Marolf and S.~F.~Ross,
  ``{\it Boundary conditions and new dualities: Vector fields in AdS/CFT},''
  JHEP {\bf 0611}, 085 (2006)
  [arXiv:hep-th/0606113].
\bibitem{guth}   E.~J.~Weinberg and A.~H.~Guth,
  ``{\it Nonexistence Of Spherically Symmetric Monopoles With Multiple Magnetic
  Charge},''
  Phys.\ Rev.\  D {\bf 14}, 1660 (1976).
\bibitem{symmon} N.~J.~Hitchin, N.~S.~Manton and M.~K.~Murray,
  ``{\it Symmetric Monopoles},''
  Nonlinearity {\bf 8}, 661 (1995)
  [arXiv:dg-ga/9503016].
\bibitem{sm2}  C.~J.~Houghton and P.~M.~Sutcliffe, ``{\it Octahedral and dodecahedral monopoles},''
  arXiv:hep-th/9601147.
\bibitem{monskyrme}  C.~J.~Houghton, N.~S.~Manton and P.~M.~Sutcliffe,
  ``{\it Rational maps, monopoles and Skyrmions},''
  Nucl.\ Phys.\  B {\bf 510}, 507 (1998)
  [arXiv:hep-th/9705151].
\bibitem{stolen2}
 R.~A.~Battye, C.~J.~Houghton and P.~M.~Sutcliffe,
 ``{\it Icosahedral Skyrmions},''
  J.\ Math.\ Phys.\  {\bf 44}, 3543 (2003)
  [arXiv:hep-th/0210147].
\bibitem{SilvaLobo:2009qw}
  J.~Silva Lobo and R.~S.~Ward,
  ``{\it Skyrmion Multi-Walls,}''
  J.\ Phys.\ A  {\bf 42} (2009) 482001
  [arXiv:0910.5457 [hep-th]]. 
\bibitem{private} R. S. Ward, private communication
\bibitem{spectral} S. Cherkis and R. S. Ward, to appear
\bibitem{shamit}  S.~Kachru, A.~Karch and S.~Yaida,
  ``{\it Holographic Lattices, Dimers, and Glasses},''
  Phys.\ Rev.\  D {\bf 81}, 026007 (2010)
  [arXiv:0909.2639 [hep-th]].
\bibitem{dw}   A.~Aperis, P.~Kotetes, E.~Papantonopoulos, G.~Siopsis, P.~Skamagoulis and
    G.~Varelogiannis,
  ``{\it Holographic Charge Density Waves},''
  arXiv:1009.6179 [hep-th].
\bibitem{stripe}  R.~Flauger, E.~Pajer and S.~Papanikolaou, ``{\it A Striped Holographic
    Superconductor},''
  arXiv:1010.1775 [hep-th].
\bibitem{qhs} S. L. Sondhi, A. Karlhede, S. A. Kivelson  and E. H. Rezayi, ``{\it Skyrmions and the
    crossover from the integer to fractional quantum Hall effect at small Zeeman energies}", Phys.
    Rev. B 47, 16419–16426 (1993)
\bibitem{sarma} K. Yang, S. Das Sarma, A. H. MacDonald, ``{\it Collective Modes and Skyrmion
    Excitations in Graphene SU(4) Quantum Hall Ferromagnets}", Phys. Rev. B 74, 075423 (2006)
    [arXiv:cond-mat/0605666]
\bibitem{maeda}   K.~Maeda, M.~Natsuume and T.~Okamura,
  ``{\it Vortex lattice for a holographic superconductor},''
  Phys.\ Rev.\  D {\bf 81}, 026002 (2010)
  [arXiv:0910.4475 [hep-th]].
  \bibitem{ooguri1}
  S.~Nakamura, H.~Ooguri and C.~S.~Park,
  ``{\it Gravity Dual of Spatially Modulated Phase},''
  Phys.\ Rev.\  D {\bf 81}, 044018 (2010)
  [arXiv:0911.0679 [hep-th]].
\bibitem{ooguri2}   H.~Ooguri and C.~S.~Park,
  ``{\it Holographic End-Point of Spatially Modulated Phase Transition},''
  arXiv:1007.3737 [hep-th].
  \bibitem{erickim}  K.~M.~Lee and E.~J.~Weinberg,
  ``{\it BPS Magnetic Monopole Bags},''
  Phys.\ Rev.\  D {\bf 79}, 025013 (2009)
  [arXiv:0810.4962 [hep-th]].
\bibitem{denef}   F.~Denef and S.~A.~Hartnoll,
  ``{\it Landscape of superconducting membranes},''
  Phys.\ Rev.\  D {\bf 79}, 126008 (2009)
  [arXiv:0901.1160 [hep-th]].
\bibitem{bhbag}   S.~Bolognesi, ``{\it Magnetic Bags and Black Holes},''
  arXiv:1005.4642.
\bibitem{rocha}  S.~S.~Gubser and F.~D.~Rocha, ``{\it The gravity dual to a quantum critical point
    with spontaneous symmetry breaking},''
  Phys.\ Rev.\ Lett.\  {\bf 102}, 061601 (2009)
  [arXiv:0807.1737 [hep-th]].
\bibitem{weak}   N.~Arkani-Hamed, L.~Motl, A.~Nicolis and C.~Vafa,
  ``{\it The string landscape, black holes and gravity as the weakest force},''
  JHEP {\bf 0706}, 060 (2007)
  [arXiv:hep-th/0601001].






\end{thebibliography}
\end{document}